\documentclass[11pt]{article}

\usepackage[utf8]{inputenc}
\usepackage[T1]{fontenc}

\usepackage{arxiv}

\usepackage{amsmath}
\usepackage{amssymb}
\usepackage{amsfonts}
\usepackage{graphicx}
\usepackage{float}
\usepackage{booktabs}
\usepackage{subcaption}
\usepackage{multirow}
\usepackage{array}
\usepackage{enumitem}
\usepackage{nicefrac}
\usepackage{tcolorbox}

\usepackage{hyperref}
\usepackage{url}
\hypersetup{
  colorlinks=true,
  linkcolor=arxivaccent,
  citecolor=arxivaccent,
  urlcolor=arxivaccent,
  filecolor=arxivaccent,
  breaklinks=true,
}

\newtcolorbox{examplebox}{colback=arxivaccent!4!white,colframe=arxivaccent!60!black,
  boxrule=0.5pt,arc=2pt,left=6pt,right=6pt,top=5pt,bottom=5pt}

\title{Transcribe, Then Reason:\\
Two-Pass Decomposition for Multimodal Review}

\author{%
  \begin{tabular}{@{}c@{\hspace{4em}}c@{}}
    Bojie Li & Noah Shi \\
    Pine AI & University of Washington
  \end{tabular}%
}
\date{}
\runningtitle{Transcribe, Then Reason: Two-Pass Decomposition for Multimodal Review}

\begin{document}

\maketitle
\vspace{-1.2em}

\begin{abstract}
The natural way to review a long recording or document with a multimodal model is to hand it the raw source and ask for a review in one call. We show that this quietly fails: the model \emph{satisfices}, dropping roughly a third of the content and embellishing the rest. The failure is not perception---almost all of the dropped content reappears when the same model is simply asked to transcribe the source. The bottleneck is \textbf{generation under load}: a single pass cannot perceive, reason over, and write a long faithful review at the same time, because doing all three competes for one output. We rule out the obvious alternatives. It is not the modality: models read text and an image of the same text equally well. And it is not merely a matter of thinking harder: giving the single pass a far larger reasoning budget does not recover the lost content, because the model spends that budget planning a review rather than writing the source down. What works is to split the labor across two same-weights passes---first transcribe, then review the transcript---so each step gets a full output budget of its own. This \emph{transcribe-then-review} decomposition improves both faithfulness and coverage across a 21-source suite. The benefit is not uniform: we observe that it helps most where the one-pass baseline is weakest and little where that baseline is already strong, a pattern that also tracks the source's length and modality. Decomposition comes with two failure modes---the review pass running out of room on very long sources, and confabulating from memory once the grounding source is removed.
\end{abstract}
\vspace{-0.4em}
\begin{center}
\footnotesize
Code: \url{https://github.com/19PINE-AI/cascade-gap} \quad$\cdot$\quad Website: \url{https://01.me/research/cascade-gap}
\end{center}
\vspace{-1.2em}

\begin{figure}[H]
\centering
\includegraphics[width=\linewidth]{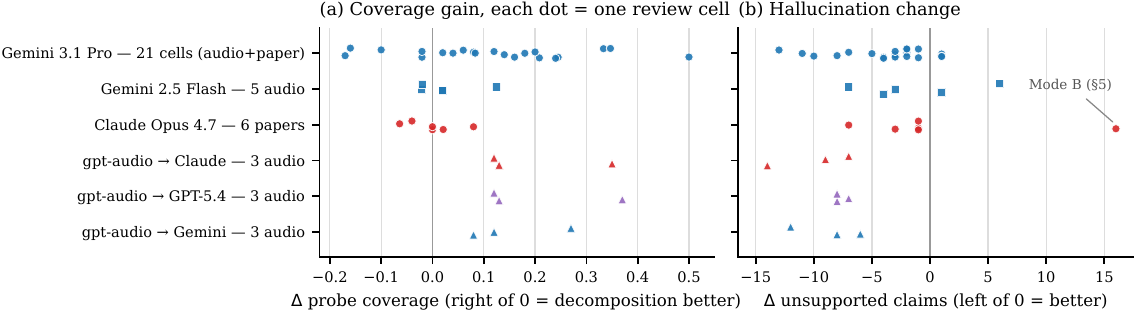}
\vspace{-8pt}
\caption{\textbf{Two-pass decomposition improves long-form review.} Each dot is one cell's change from end-to-end review to transcribe-then-review: \textbf{(a)} probe coverage and \textbf{(b)} unsupported claims. Rows use the same weights; \texttt{gpt-audio} rows are mixed pipelines.}
\label{fig:headline}
\end{figure}

\section{Introduction}
\label{sec:intro}

\begin{figure}[t]
\centering
\includegraphics[width=0.72\linewidth]{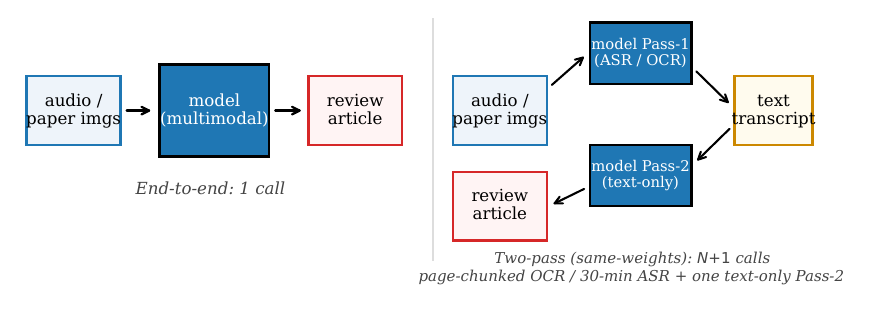}
\caption{\textbf{Two configurations with the same weights.} End-to-end review writes from raw audio or page images in one call. Transcribe-then-review transcribes first, then writes from that transcript alone; only the reasoning step's input representation differs.}
\label{fig:schematic}
\end{figure}

Multimodal models can read audio and document images directly, so the obvious way to have one review a long talk or paper is end-to-end: hand it the raw source and ask for a review in a single call (Figure~\ref{fig:schematic}, top). This feels like the safe choice. Transcribing first can only lose information, the reasoning goes, whereas an end-to-end model gets to see and hear everything. We find the intuition is backwards, and the rest of the paper explains why.

Asked to review a long source in one pass, frontier models \emph{satisfice}: they produce a fluent review that silently omits much of the content and invents some of the rest. On our running example---Andrej Karpathy's 42-minute \emph{State of GPT} talk---Gemini~3.1~Pro's end-to-end review covers only about six in ten of the facts an independent judge extracts, and roughly a quarter of its claims are unsupported by the source (Figure~\ref{fig:example}). The natural guess is that the model could not hear or read the source well enough. That guess is wrong: when we ask the \emph{same model} to simply transcribe the same audio, its transcript contains essentially every fact the review dropped---in the running example, the missing sentences appear word for word. The content is perceived; it is lost somewhere between perceiving it and writing the review.

The loss is a \emph{generation} problem, not a perception or modality problem. We test the two competing explanations and both fail. It is not that models reason worse over pixels or audio than over text: they read the same facts equally well from text and from an \emph{image} of that text, and, under reasoning load, sometimes better from the image (\S\ref{sec:mechanism}). And it is not simply that the model needs to think longer: giving the single pass a much larger reasoning budget does not recover the lost content, because the model uses that budget to \emph{plan} a review, not to write the source down for itself. What actually helps is giving the model a second output: when one call is forced to both transcribe and review, it either compresses the review to nothing or runs out of room, but two separate calls---each with its own full output budget---do not. The bottleneck is that a single pass cannot perceive, reason, and write a long faithful review all at once; the three jobs compete for one act of generation.

This points to a simple fix: give each job its own pass, using the same model throughout. Pass~1 transcribes the source---\emph{perceive and externalize}. Pass~2 writes the review from that transcript alone---\emph{synthesize}. This transcribe-then-review decomposition improves faithfulness and coverage over the end-to-end baseline across our 21-source suite (Figure~\ref{fig:headline}). Crucially, it is not a uniform win, and it is not supposed to be. It helps in proportion to how badly the one-pass baseline was satisficing---most where the baseline is weak, little where it is already strong. This same observation (\S\ref{sec:solution}) turns a messy set of cross-vendor results into one coherent picture (Figure~\ref{fig:xmodel}), and it points to the two situations in which decomposition breaks: the review pass running out of room on a very long source, and the review pass confabulating from memory once the original source is taken away (\S\ref{sec:failure_modes}).

\paragraph{Contributions.}
(1)~We document that end-to-end long-form review satisfices, and show through a perception check that the loss is downstream of perception (\S\ref{sec:observation}).
(2)~We identify the cause as generation under load, ruling out the modality and ``think-longer'' explanations with same-weights controls including behavioral probes on the frontier models (\S\ref{sec:mechanism}).
(3)~We show the transcribe-then-review fix and characterize when it helps---an observation that ties together baseline headroom, modality, and two failure modes---across two modalities, three vendors, and mixed pipelines that let text-only models review audio.

\section{The observation: end-to-end long-form review satisfices}
\label{sec:observation}

\begin{figure}[!t]
\begin{examplebox}
\small
\textbf{Running example: Gemini 3.1 Pro reviews Karpathy's \emph{State of GPT} (42-min audio), end-to-end.}
All excerpts are verbatim from the released run artifacts (Appendix~\ref{appdx:code}).\\[4pt]
\textbf{The talk says:} \emph{``The vocabulary size is usually a couple ten thousand tokens''} --- and the model hears it: its own transcript of the audio contains the sentence verbatim.\\[2pt]
\textbf{The end-to-end review writes:} \emph{``A typical vocabulary size is around 10{,}000 tokens.''} --- one of \textbf{13 unsupported claims}, alongside a misdescription of RLHF as a fourth serial pipeline stage.\\[4pt]
\textbf{The talk ends with} Karpathy prompting GPT-4: \emph{``can you say something to inspire the audience of Microsoft Build 2023?''} --- again present verbatim in the model's own transcript.\\[2pt]
\textbf{The end-to-end review:} drops it entirely, one of \textbf{19 of 51 probes missed}. All 19 missed probes are present in the model's own 8{,}500-word transcript of the same audio.\\[4pt]
\textbf{The two-pass review} (\S\ref{sec:solution}: same weights, written from that transcript alone) covers \textbf{49 of 51 probes with 3 unsupported claims}, and gets the example right: \emph{``Vocabulary sizes are usually a couple ten thousand tokens (e.g., 50{,}257).''}
\end{examplebox}
\vspace{-2pt}
\caption{\textbf{Running example.} The end-to-end review drops perceived content and adds unsupported detail; the two-pass review recovers coverage and faithfulness. Karpathy is the bolded row in Figure~\ref{fig:headline_cells}a.}
\label{fig:example}
\end{figure}

\paragraph{Setup.}
The task is long-form review: given a recording (26--85 minutes) or a paper supplied as page images, write a review that is both \emph{exhaustive} (covers what the source says) and \emph{faithful} (adds nothing the source does not say). To measure both, an independent judge (GPT-5.4) extracts about fifty atomic factual ``probes'' from a reference transcript of each source, marks each probe as covered or missing in a given review, and separately lists the review's unsupported claims. Our suite is 21 cells: 11 audio cells from 10 recordings---talks, lectures, panels, and meetings (Karpathy yields both a review and a meeting-minutes task)---and 10 papers, ranging from 1968--1972 NASA reports to post-cutoff 2026 arXiv reviews. The measurement details, noise thresholds, and confidence intervals are in Appendix~\ref{appdx:protocol} and~\ref{appdx:cell_table}. Throughout, \emph{end-to-end review} is the one-call configuration: the model reads the raw source and writes the review in a single call.

\paragraph{End-to-end review omits roughly a third of the content.}
Across the suite, the end-to-end review covers on average only about two-thirds of the probes, and every cell contains unsupported claims, or \emph{hallucinations} as the tables label them (Figure~\ref{fig:headline_cells}, gray points). The running example (Figure~\ref{fig:example}) shows the three characteristic errors in miniature: invented precision (``10{,}000 tokens'' for what the speaker calls ``a couple ten thousand''), a structural mistake (describing RLHF as a fourth pipeline stage), and outright omission (the talk's closing demonstration simply vanishes).

\begin{figure}[t]
\centering
\includegraphics[width=\linewidth]{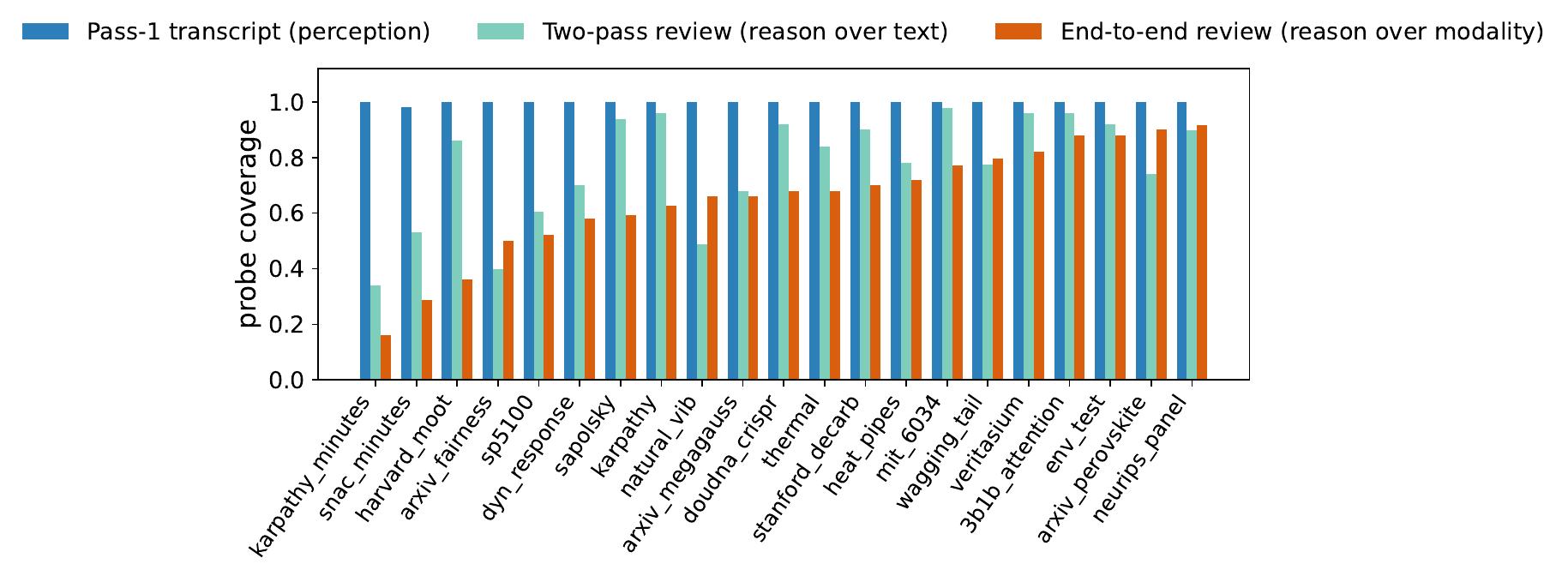}
\caption{\textbf{Perception check.} Per cell, the model's own transcript (blue) covers nearly all probes---far more than either the end-to-end (orange) or transcribe-then-review (green) review. Almost all content the end-to-end pass drops is present in that transcript.}
\label{fig:mech_e3}
\end{figure}

\paragraph{The dropped content was perceived.}
The obvious explanation is that the model could not hear or read the source well enough. It could. When we ask the same model to \emph{transcribe} each source instead of review it, the transcript covers essentially every probe---far more than the review the model wrote from the very same input (Figure~\ref{fig:mech_e3}). Almost none of the omitted content is genuinely lost: $99.7\%$ of what the end-to-end review drops is present, as text, in the model's own transcript (in the running example, all 19 dropped facts). The transcript's near-perfect coverage is partly by construction, since the probes come from a similar transcript; the point we need is narrower and robust to that---the dropped content \emph{was} within the model's reach as text. So the loss happens after perception, when the model writes the review. The next section asks why.

\section{Mechanism: why a single pass fails}
\label{sec:mechanism}

The model perceives the content and then loses it while writing. Why? There are three natural explanations, and they make different predictions, so we can tell them apart. The \textbf{modality hypothesis} says the model reasons better over text than over raw audio or pixels, so writing directly from the modality is handicapped. The \textbf{generation-load hypothesis} says the problem is having one pass do everything at once---perceive, reason, and write a long faithful review---so that the three jobs compete for a single output. The \textbf{attention-dilution hypothesis} says the long, low-information stream of audio or image tokens simply crowds out the model's attention. The rest of this section rules out the first, cannot test the third on our frontier model, and lands on the second. Full per-experiment tables are in Appendix~\ref{appdx:substrate}.

\begin{figure}[t]
\centering
\begin{subfigure}{\linewidth}
\centering
\includegraphics[width=0.88\linewidth]{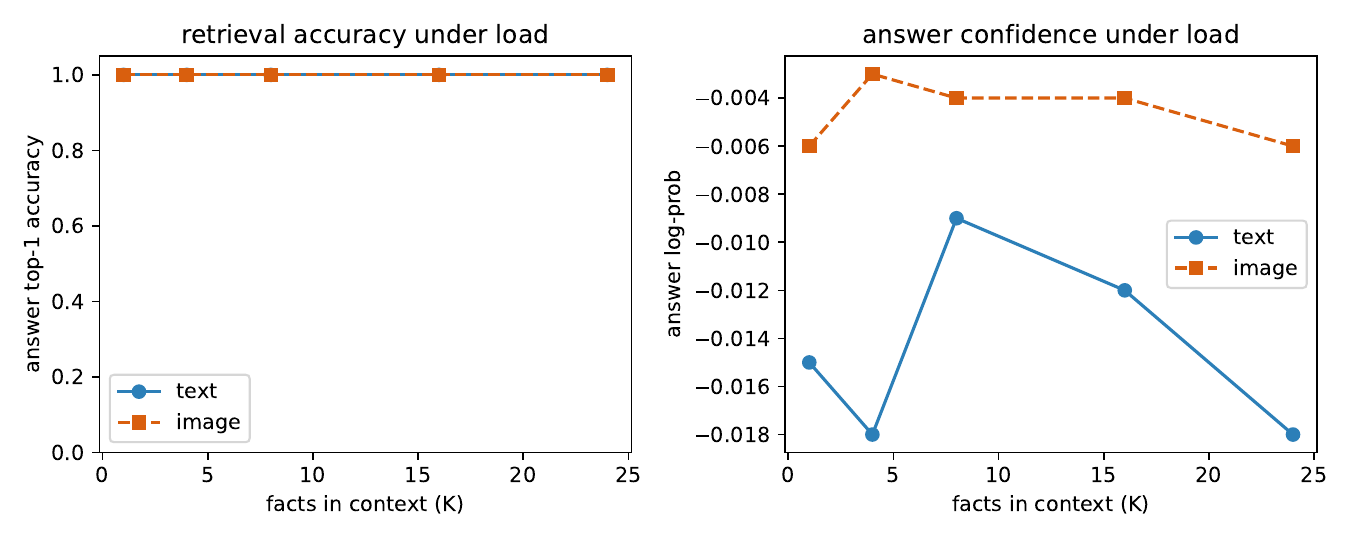}
\caption{Vision activation probe (open VLM): layer-wise logit-lens on Qwen2.5-VL-7B.}
\end{subfigure}\\[4pt]
\begin{subfigure}{\linewidth}
\centering
\includegraphics[width=\linewidth]{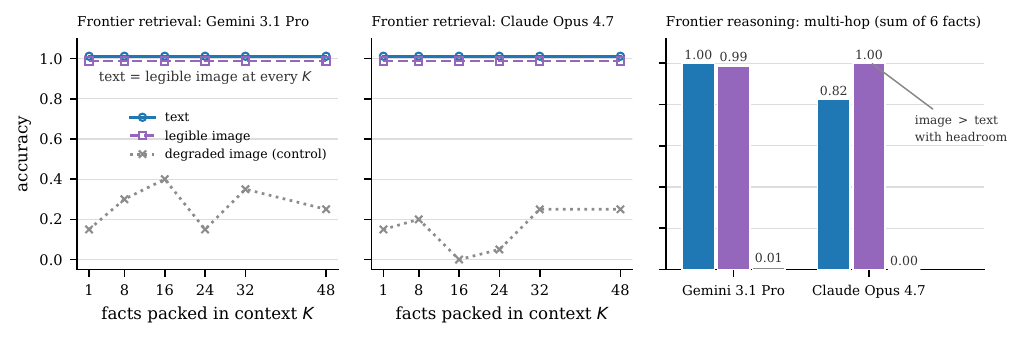}
\caption{Frontier probes, behavioral (the headline frontier models).}
\end{subfigure}
\caption{\textbf{Legible pixels are not a worse reasoning medium than text.} \textbf{(a)} Open-VLM activation probes find equivalent text/image readout under multi-fact load; the intelligible-audio analogue is also null. \textbf{(b)} On the headline models, retrieval from text and legible images is identical, while a degraded-image control confirms the probe detects real deficits; in multi-hop reasoning the one model with headroom (Claude) does no worse from images---indeed better.}
\label{fig:substrate}
\end{figure}

\paragraph{The modality is not the bottleneck.}
If pixels or audio were a worse reasoning medium than text, the fix would be to convert to text---and the modality, not the number of passes, would be doing the work. But when we hold the content fixed and vary only its \emph{form}, the penalty does not appear (Figure~\ref{fig:substrate}). We render the same novel facts as plain text and as an \emph{image} of that text and ask the model to read them back. Small open models do this equally well from either form, from one fact up to two dozen packed together; the frontier models do too, right up to 48 facts on a page, while a deliberately blurred-image control confirms the test can detect a real deficit when there is one. The audio analogue is the same story: an apparent gap under robotic text-to-speech disappears once the voice is a clean neural one. And when the task requires actual reasoning rather than retrieval---pulling several facts off the page and combining them---the one frontier model with room to fail does \emph{better} from the image than from the text. The scope is legible propositional content, not diagrams or 2D layout; within it, there is no general penalty for reasoning over pixels. The contrast with the review result is stark: these same models read individual facts off an image perfectly, yet drop a third of the content when they review. Reading the modality is not the problem.

\begin{figure}[t]
\centering
\includegraphics[width=0.82\linewidth]{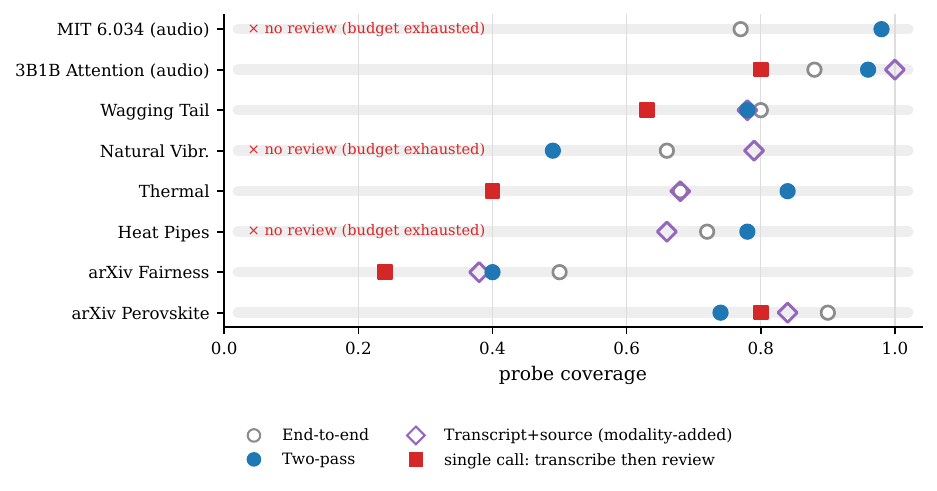}
\caption{\textbf{The benefit requires two output budgets, not removal of the modality.} On the 8-cell grid, a one-call ``transcribe then review'' prompt (red squares) usually underperforms transcribe-then-review and sometimes emits no review. Adding the raw modality back to Pass~2 (the transcript-plus-source control; diamonds) does not restore the coverage benefit and adds unsupported claims.}
\label{fig:e2_collapse}
\end{figure}

\paragraph{The bottleneck is generation under load.}
That leaves the load explanation, and two controls pin it down. The first asks whether the fix is just getting the text into the context window. If it were, a single call told to transcribe verbatim \emph{and then} review should do as well as two separate calls. It does not (Figure~\ref{fig:e2_collapse}): forced to do both in one response, the model either shrinks the review far below the two-pass level or spends its whole output on the transcript and never reaches the review. What matters is not that the transcript exists somewhere, but that the review gets its own act of generation with a full budget behind it. The second control rules out the reverse worry---that the raw pixels or audio in context are actively poisoning the review. Handing the review pass \emph{both} the transcript and the original modality does not help coverage and, if anything, brings back some unsupported claims relative to working from text alone. Removing the modality is therefore a minor faithfulness convenience; the real ingredient is the split into two passes, each with its own output budget.

\paragraph{Additional reasoning budget does not recover the loss.}
A tempting objection is that this is trivial: two passes just give the model more total budget, so surely one pass would recover if we let it think longer. It does not. Holding the end-to-end configuration fixed and sweeping only the model's reasoning budget across its full supported range (a $256\times$ span) does not lift coverage toward the two-pass level: on our eight-cell subset the maximum-budget coverage equals the default-budget coverage within noise (mean difference $+0.005$), and on the cells where two-pass genuinely helps it closes almost none of the gap---on MIT~6.034, one-pass covers $0.77$ and two-pass $0.98$, yet even the maximum reasoning budget reaches only $0.81$ (Appendix~\ref{appdx:substrate}). Inspecting what the model does with the extra budget shows why: it spends the reasoning on \emph{planning} the review---organizing the talk into sections, deciding what to emphasize---not on writing the source down for itself. The transcript never appears in its reasoning. More thinking buys a better-planned but still satisficed review, because the missing step is externalizing the content into text, and the model will not spend a single generation both externalizing and reviewing (which is exactly what the forced single-call control already showed). The recovery needs a \emph{second} generation, not a longer first one.

\section{The solution: two-pass decomposition, and where it wins}
\label{sec:solution}

\paragraph{The configuration.}
The fix follows directly from the mechanism (Figure~\ref{fig:schematic}). In transcribe-then-review, Pass~1 transcribes the source to text and Pass~2 writes the review from that transcript alone. Same model, same weights; the only change from end-to-end review is that the writing step now works from text it produced, each step with a full output budget. Later sections introduce small variants of Pass~2 as we need them---chunking it for very long sources, or re-grounding it in the transcript---but plain transcribe-then-review is the object of interest here.

\begin{figure}[t]
\centering
\includegraphics[width=\linewidth]{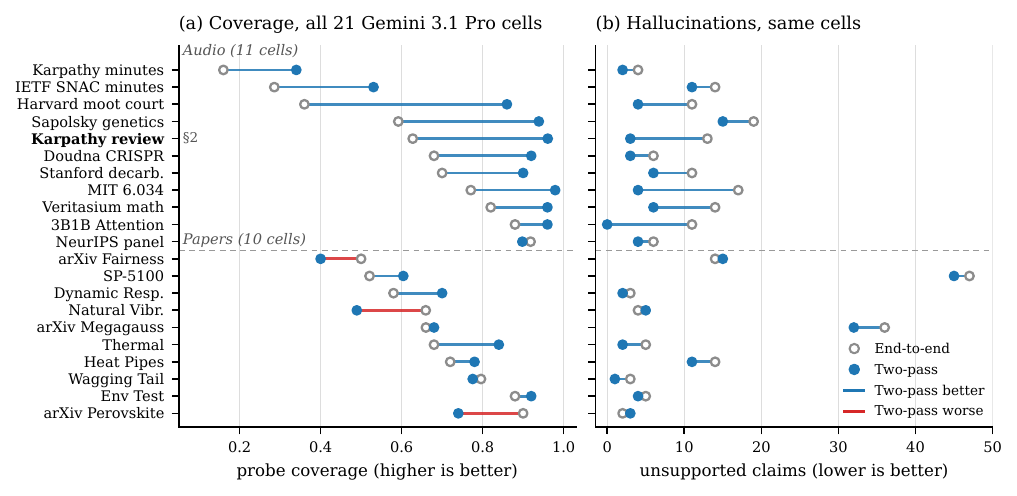}
\caption{\textbf{Per-cell results for all 21 Gemini~3.1~Pro cells.} Cells are sorted by end-to-end coverage within modality; the bold row is the running example. Transcribe-then-review improves coverage and reduces unsupported claims on most cells; red connectors mark regressions.}
\label{fig:headline_cells}
\end{figure}

\paragraph{Headline result.}
Across the 21 sources, transcribe-then-review helps on both axes: it reduces unsupported claims on 18 of 21 cells (mean $-3.86$, 95\% CI $[-5.62,-2.33]$) and improves coverage on 16 of 21 (mean $\Delta_{\rm cov}=+0.118$, 95\% CI $[+0.048,+0.189]$); both effects are sign-test significant ($p=0.0015$ and $p=0.027$; per-cell numbers and confidence intervals in Figure~\ref{fig:headline_cells} and Appendix~\ref{appdx:cell_table}). The gains are largest exactly where end-to-end review was worst---Harvard Moot Court, the weakest baseline, gains the most coverage, and 3Blue1Brown's Attention lecture goes from eleven unsupported claims to zero. The direction survives the robustness checks of \S\ref{sec:robust}. One caveat carries through the paper: Pass~1 and the reference transcript come from the same model, which could flatter transcribe-then-review. Regenerating the reference with an independent transcription system preserves the direction on all four cells we could test, but that control is small, so we treat this shared-model bias as the main residual threat to the headline number.

\begin{figure}[t]
\centering
\includegraphics[width=0.62\linewidth]{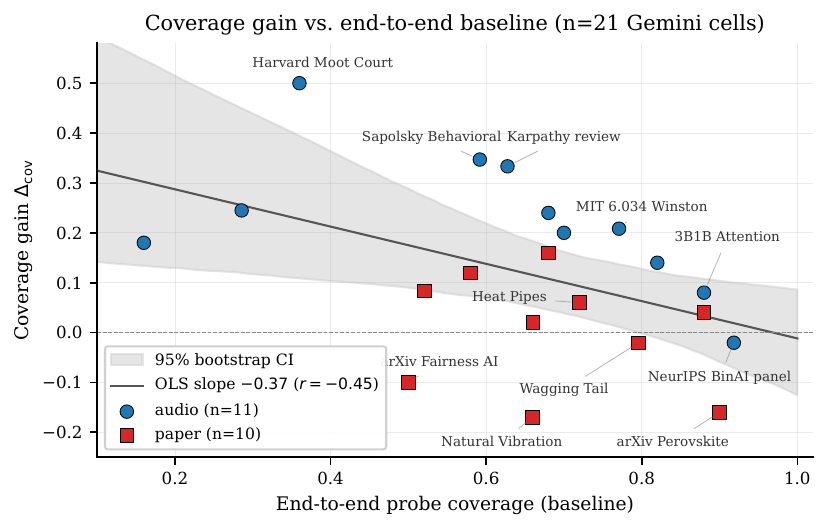}
\caption{\textbf{Coverage gain versus end-to-end baseline.} Cells with a weaker end-to-end baseline tend to gain more from decomposition, most clearly among the audio cells; circles denote audio and squares papers. The line and band are an ordinary least-squares (OLS) fit and its bootstrap CI. Cells that gain little either start near the ceiling or fall into the two failure modes (Modes~A/B).}
\label{fig:solution}
\end{figure}

\paragraph{Where decomposition helps.}
Decomposition does not help uniformly, and the pattern is legible in Figure~\ref{fig:solution}: the gains are largest where the end-to-end baseline is weakest and shrink as it improves ($r=-0.45$, bootstrap 95\% CI $[-0.73,-0.10]$). This is what the mechanism predicts---a low baseline means the one-pass model left perceived content unwritten, and the perception check showed that content is sitting in the transcript, within reach of a full-budget synthesis pass; where the baseline is already high there is little headroom, so the second pass changes little. The figure also shows that this is not one clean relationship: the trend is clearest among the audio cells, while the papers cluster at smaller gains because they are longer and more citation-dense, and so are the ones that hit the two failure modes of \S\ref{sec:failure_modes}. We read Figure~\ref{fig:solution} as an observation in which modality, baseline headroom, and the failure modes are entangled rather than as a clean rule; the cells where transcribe-then-review does not help are not counterexamples but either near-ceiling or failure-mode cases. The practical upshot is still simple: spend the second pass where the one-pass model is under-performing.

\begin{figure}[t]
\centering
\includegraphics[width=0.98\linewidth]{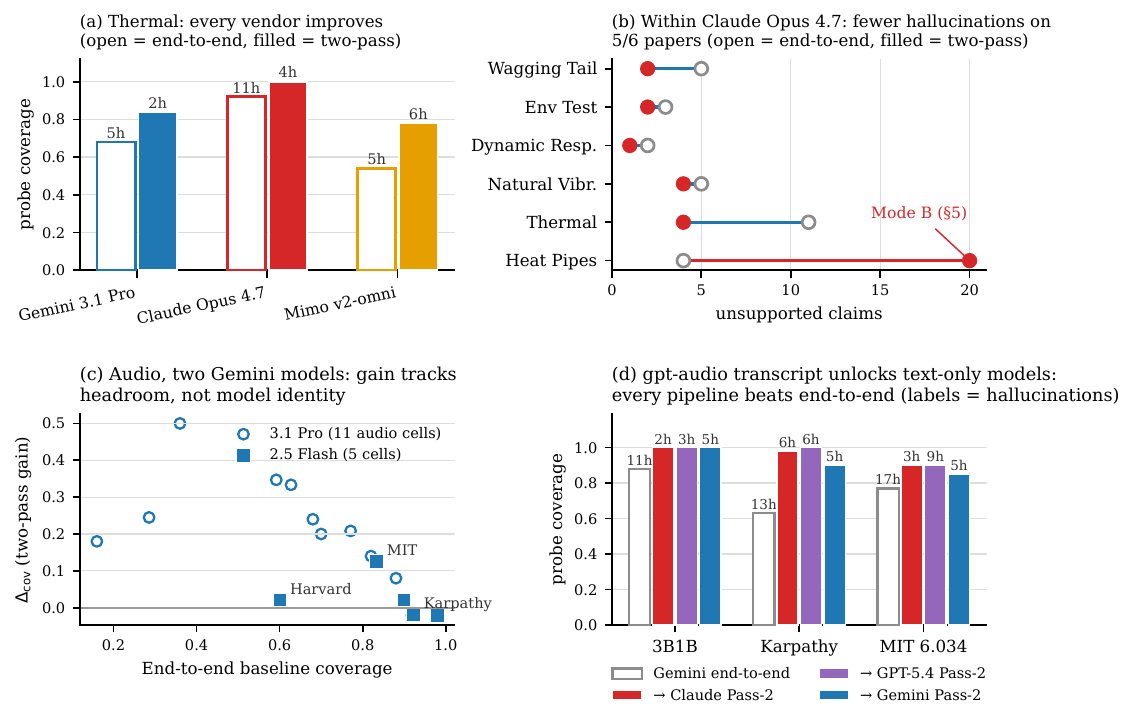}
\caption{\textbf{Cross-model results track baseline headroom rather than replicating uniformly.} \textbf{(a)} On Thermal, all three vendors gain coverage. \textbf{(b)} In Claude's feasible NASA cells, hallucinations fall on all but Heat Pipes (Mode~B), while short papers have little coverage headroom. \textbf{(c)} Flash's strong audio baseline leaves gains only on MIT and Harvard; Karpathy regresses. \textbf{(d)} Mixed pipelines let text-only models review audio and beat the native end-to-end baseline on both axes.}
\label{fig:xmodel}
\end{figure}

\paragraph{Cross-vendor and mixed-model results.}
Run the same comparison on other models and the results look scattered: big wins on some cells, nothing on others, an occasional small regression. The same headroom pattern makes them coherent (Figure~\ref{fig:xmodel}). Gemini~2.5~Flash already reviews audio well end-to-end, so decomposition mostly washes out; Claude's short NASA papers start near the ceiling, so likewise; Gemini~Pro's weaker audio has room, so it gains. The apparent inconsistency across vendors is the same pattern---each model gains in proportion to its own headroom. Two honest limits: our headline aggregate is single-vendor (Gemini), and a clean cross-\emph{company} audio cascade is impossible today, because no non-Google frontier model both ingests audio and accepts a text-only second pass (\S\ref{sec:discussion}). That constraint is why the most deployment-relevant result is the mixed pipeline (Figure~\ref{fig:xmodel}d): transcription lets a text-only frontier model review audio it could not otherwise touch, beating the native end-to-end baseline on both axes (per-vendor detail in Appendix~\ref{appdx:xvendor}; the mixed-pipeline and full-quality Gemini~2.5~Flash arms in Appendices~\ref{appdx:mixed} and~\ref{appdx:flash_audio}).

\paragraph{Cost.}
Decomposition is not free: it replaces one call with $N{+}1$ (one transcription pass---chunked for long sources---plus the review), so it costs more where the source is longer. Because the gains concentrate on weak-baseline cells and vanish on saturated ones, the same headroom pattern is also a useful cost guide: spend the extra calls where end-to-end has headroom, not where it is already strong. Table~\ref{tab:checklist} turns this into a per-regime deployment checklist, and the full cost-versus-gain picture is in Appendix~\ref{appdx:cell_table}.

\section{Where decomposition breaks: two failure modes and their fixes}
\label{sec:failure_modes}

\begin{figure}[t]
\centering
\includegraphics[width=0.82\linewidth]{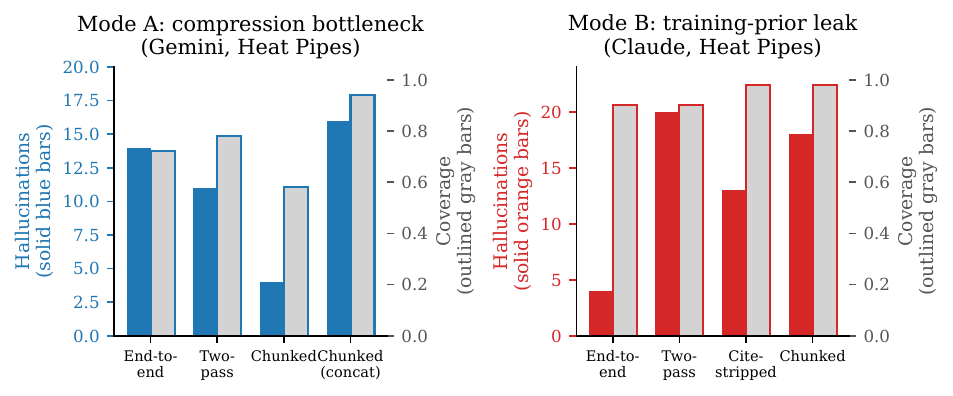}
\caption{\textbf{The two failure modes, on the Heat Pipes paper.} Gemini's cascade hits mostly Mode~A (a synthesis-budget bottleneck, relieved by chunking); Claude's hits mostly Mode~B (training-prior confabulation). Intervention results are in Appendix~\ref{appdx:fixes}.}
\label{fig:failure_modes}
\end{figure}

The mechanism does not just predict that decomposition helps; it predicts the two specific ways it can break, and we see both. Each is a seam in the two-pass split. \textbf{Mode~A} is a \emph{room} problem: the review pass has a bounded output (a couple of thousand words), so when the transcript is very long---a long paper---the review physically cannot hold everything and drops content. \textbf{Mode~B} is a \emph{grounding} problem: because Pass~2 sees only the transcript and not the original source, it loses the anchor that would stop it from filling in details from memory. On a paper that cites famous prior work, the model quietly expands what the transcript says using what it already knows---adding facts that are plausible, often correct, but not in the source and therefore unfaithful. Figure~\ref{fig:failure_modes} shows both, on the same paper (Heat~Pipes), with Gemini hitting mostly Mode~A and Claude mostly Mode~B.

\paragraph{Mode~A: the synthesis-budget bottleneck.}
Mode~A shows up on the longest papers, where the transcript is many times longer than the review pass can output. The fix is mechanical: instead of asking one review pass to compress the whole transcript, \textbf{chunk} it---split the transcript into sections, review each, and concatenate. This gives the content more total room and recovers most of the lost coverage; on the Heat~Pipes paper it raises coverage from $0.78$ to $0.94$ (Figure~\ref{fig:failure_modes}). The remedy replicates on two of the three long arXiv papers we tried it on, at the cost of a few added unsupported claims where the chunk boundaries lose context (details in Appendix~\ref{appdx:fixes}).

\paragraph{Mode~B: training-prior confabulation.}
Mode~B is best seen in a single example. The Heat~Pipes paper refers to ``Reference~2 (Cotter)''; working from the transcript alone, Claude expands this to ``Cotter's 1965 analysis (LA-3246-MS)''---a real citation, but the year and report number come from the model's training, not from the source (more examples in Appendix~\ref{appdx:examples}). This bibliographic-expansion signature appears on exactly the papers the mechanism predicts: long, citation-dense documents whose references are well-represented in pretraining, and not on citation-light sources (Appendix~\ref{appdx:cell_table} gives the per-cell citation densities). The deployment trigger is a source dense with citations to well-known prior work; the per-1k hallucination densities that separate this effect from mere verbosity are in Appendix~\ref{appdx:perword}.

\paragraph{Interventions for the two modes.}
Two interventions target the modes (full results in Appendices~\ref{appdx:fixes} and~\ref{appdx:iter}). Chunking Pass~2 reliably relieves Mode~A on most long papers. Mode~B has no clean remedy: stripping citations from the transcript does not survive run-to-run noise, and a quote-grounded refinement---which requires each claim to carry a verbatim transcript quote---suppresses confabulation only where Mode~B is the binding constraint, trading coverage away elsewhere.

\paragraph{Robustness of the main result.}
\label{sec:robust}
We test the headline effect against the measurement threats that most concern us: a single judge's idiosyncrasies, run-to-run randomness in the review pass and the judge, circularity from one model both writing and scoring the probes, and shared-model reference bias. The effect holds under all four: a second judge agrees on most cells, the wins survive five re-runs, re-extracting the probes with a different model preserves the coverage direction, and an independent transcription reference preserves it on all four cells we could test. That last check is small and concentrated on strong-win cases, so shared-model reference bias remains the main residual threat to the headline number. One methodological point carries into how the rest of the paper should be read: probe coverage is far more stable than the unsupported-claim count, which can swing on the few cells with very long claim lists---so we rely on coverage for the main conclusions and treat large hallucination counts as uncertain. The checks in full, with their figures and per-source numbers, are in Appendix~\ref{appdx:robustness} and~\ref{appdx:stoch}.

\section{Related work}
\label{sec:related}

\paragraph{Speech and documents.} Prior cascade-vs-end-to-end results are often read as modality claims; our mechanism study suggests they are at least partly about decomposition and modality-specific recoverability. Our Pass~1 instantiates standard transcription---large-scale weakly-supervised ASR~\cite{radford2022whisper} for audio and neural document OCR~\cite{blecher2023nougat} for pages---so the contribution is the two-pass split, not the transcriber. Cuervo et al.~\cite{cuervo2025closing} document a text/speech understanding gap, and the Cascade Equivalence Hypothesis~\cite{ceh2026} shows speech-LLMs \emph{internally} behave like ASR$\to$LLM cascades, while X-Talk~\cite{xtalk2025} finds explicitly cascaded speech-to-speech pipelines competitive. On documents, DocVLM~\cite{docvlm2024} reports OCR-augmented VLMs beating pure-vision VLMs on DocVQA at fixed budget, while the OCR-Reasoning Benchmark~\cite{ocrreasoning2026} shows pure-OCR cascades cap below 50\% where spatial layout carries meaning, ColPali~\cite{colpali2025} shows vision-only retrieval beating OCR, and GUI agents do better from raw pixels than from structured-text intermediates~\cite{uground2024,uitars2025}. A parallel vision-side literature isolates the modality by rendering identical content as text versus as an \emph{image} of that text: VISTA-Bench~\cite{vistabench2026} finds a pronounced gap across 20+ open VLMs but a near-zero gap on the strongest models---including Qwen2.5-VL ($-2.0$ points), the model in our vision activation probe---consistent with our modality null (\S\ref{sec:mechanism}), while CrossMath~\cite{crossmath2026} bounds it: on 2D symbolic-arithmetic grids, reasoning from an \emph{image} collapses relative to a markdown-table rendering of the same grid ($12.4$ vs.\ $92.8$ on Qwen3.5-Plus) with OCR intact. Its authors attribute this to a modality deficit; we read the same numbers---and the fact that serializing the identical layout to text restores reasoning---as a penalty specific to recovering 2D structure from pixels rather than to legible content per se. The common thread matches our mechanism: decomposition helps when the task is generation-under-load and the content survives transcription, and hurts when the answer lives in layout or signal that transcription discards.

\begin{table}[!t]
\small
\centering
\caption{\textbf{Practitioner checklist} drawn from this paper's cells: which configuration to deploy per source/model regime.}
\label{tab:checklist}
\begin{tabular}{p{0.34\linewidth}p{0.30\linewidth}p{0.27\linewidth}}
\toprule
Source / model class & Diagnosis & Recommended action \\
\midrule
Audio talk or moderate paper on a capable model & Cascade strongly Pareto-positive. & Transcribe-then-review. \\
Moderate paper with Mode-B Gemini signature (Natural~Vibration) & Mode~B dominant, and the cell flips to Pareto-positive under quote-grounded refinement. & Quote-grounded refinement. \\
Long paper, Mode~A dominant (Heat~Pipes, 104~pages, Gemini) & Compression bottleneck. & Chunked review (no merge). \\
Long paper, Mode~B dominant (Heat~Pipes Claude) & Citation-prior leak, but the quote-grounded-refinement faithfulness gain washes out at $n{=}5$. & End-to-end review if it fits; quote-grounded refinement buys only a coverage loss. \\
Long survey (arXiv Fairness~AI, Megagauss; NASA SP-5100) & Mode~A + Mode~B compound. At $n{=}5$, quote-grounded refinement is net-negative on Fairness~AI but a faithfulness/coverage trade on SP-5100/Megagauss. Chunked review partially replicates. & Prefer end-to-end review when it fits; if a cascade is required, use chunked review with multi-seed averaging. \\
\bottomrule
\end{tabular}
\end{table}

\paragraph{Step-Audio-R1.} Step-Audio-R1~\cite{stepaudior1} treats audio LLMs' reliance on a \emph{textual surrogate} of audio as a failure to correct, but its tasks (music, environmental sound) have \emph{acoustic} answers that transcription discards; ours is the complementary regime, where the payload is what was \emph{said}, and our Mode~B is the within-regime echo of its concern.

\paragraph{Task decomposition and long context.} Breaking a hard task into steps is a well-established way to improve LLM reasoning---chain-of-thought~\cite{wei2022cot}, least-to-most~\cite{zhou2022ltm}, and decomposed prompting~\cite{khot2022decomp}. Our transcribe-then-review split is a deployment-time instance of the same idea for long-form perception, and the single-pass degradation we observe is consistent with the long-context position effects documented by \emph{lost-in-the-middle}~\cite{liu2024lost}. Our evaluation follows the faithfulness-in-summarization line~\cite{maynez2020faithful} and atomic factual-precision scoring~\cite{min2023factscore}, adjudicated by an LLM judge~\cite{zheng2023judge}.

\paragraph{Cascade theory and faithfulness.} Bengio's Consciousness Prior~\cite{bengio2017consciousness} argues for sparse discrete bottlenecks between perception and reasoning; our externalized text bottleneck is a deployment-time analogue, and Discrete JEPA~\cite{discretejepa2025} makes a related architectural argument within vision. Mode~B is a self-cascade analogue of the unfaithfulness problem motivating RAG~\cite{rag2020}, RARR~\cite{rarr2023}, and Attribution to Identified Sources~\cite{ais2022}; our quote-grounded refinement adopts RARR's in-context-attribution discipline but operates entirely within a same-weights self-cascade without external retrieval.

\section{Limitations}
\label{sec:discussion}

\begin{itemize}[leftmargin=1.3em,itemsep=3pt,topsep=2pt,parsep=0pt]
\item \textbf{Single-vendor headline.} The headline aggregate is Gemini-only. A cross-company same-weights \emph{audio} cascade is precluded by API constraints---no non-Google frontier model both ingests audio and accepts a text-only Pass-2---which is why the mixed pipeline is the natural production form.
\item \textbf{Shared-model reference.} Pass-1 and the reference transcript come from one model, the principal residual threat to the headline statistic; the independent-reference control that addresses it is small (four cells).
\item \textbf{Judge noise.} The unsupported-claim count is stochastic on high-count cells, so we rely on the more stable coverage measure for the main conclusions.
\item \textbf{Long-form review only.} We make no claim about short-form QA, code, or agentic tasks; the account predicts \emph{no} decomposition benefit where the one-pass baseline is not already satisficing.
\item \textbf{Behavioral frontier probes.} Frontier activations are inaccessible, so those probes are behavioral rather than activation-level, leaving Claude as the only headroom case on the reasoning probe.
\item \textbf{Clean pages only.} Real scanned or degraded pages---where a modality or attention-dilution effect is most likely to appear---remain untested.
\end{itemize}

\section{Conclusion}
\label{sec:conclusion}

End-to-end long-form review quietly satisfices: the model omits about a third of content it could transcribe and invents some of the rest. The bottleneck is generation under load---one pass cannot perceive, reason, and write a long faithful review at once---not the modality, and not a shortage of thinking. Giving each job its own same-weights pass, transcribe then review, recovers faithfulness and coverage; the gain concentrates where the one-pass baseline is weakest and comes with two predictable failure modes, the review pass running out of room or confabulating once the source is gone. The practical default, whenever a model struggles with a raw source, is \textbf{transcribe, then reason}.

\section*{Acknowledgements}

This paper was produced using Pine Copilot's voice-directed \emph{whisper coding} workflow~\cite{pineai2026whispercoding}, in which the authors specify, discuss, and review the work by voice while a coding agent---Claude Code with Claude Opus 4.8---carries out the planning, coding, experiments, and paper writing.
We thank BSQL Networking for hosting the NVIDIA RTX PRO 6000 GPU.

\bibliographystyle{plainnat}

\appendix

\noindent\textbf{Notation.} For compactness the tables in this appendix use short labels for the configurations named descriptively in the main text: $C_0$~=~end-to-end review (one call); $C_1$~=~transcribe-then-review (two passes); $C_{1c}$~=~chunked review; $C_1^{\rm strip}$~=~citation-stripped review; $C_{1\rm iter}$~=~quote-grounded refinement; $C_{1+}$~=~transcript-plus-source (the modality-added control).

\section{Mechanistic experiments and controls, in detail}
\label{appdx:substrate}

This appendix expands the mechanism sections (the perception check of \S\ref{sec:observation}, and the decomposition controls, modality probes, and confounded controls of \S\ref{sec:mechanism}) with the per-experiment tables.

\paragraph{Perception check: perception is not the bottleneck.}
The per-cell breakdown is Figure~\ref{fig:mech_e3} (main text). The transcript covers a mean $99.9\%$ of probes, versus $65\%$ for the $C_0$ review and $77\%$ for $C_1$ synthesis; $361/362$ probes dropped by $C_0$ are present in the model's own transcript.

\paragraph{Vision activation probe: no cross-modal gap at the activation level.}
We tested the modality hypothesis directly on an open VLM (Qwen2.5-VL-7B), controlling for prior knowledge with novel synthetic facts about nonsense entities (e.g.\ ``the Zorblatt device weighs 7 kilograms''). Each fact is presented either as digital text or as an \emph{image} of that same text, with a matched single-token question, and a layer-wise logit-lens reads off the log-probability and rank of the answer token at every layer. The two modalities are indistinguishable for an isolated fact and under load: packing $K\in\{1,4,8,16,24\}$ facts into one stimulus, image and text stay at $1.00$ accuracy with a readout-depth gap $<1$ layer (equivalent within $\pm 1$ layer under a two one-sided tests procedure, TOST) at every $K$ (Figure~\ref{fig:substrate}a). The same null replicates on Qwen2.5-VL-3B; the per-$K$ ladder for both models is in Appendix~\ref{appdx:vision_ladder}. (The 32B variant verbalizes answers in a form our single-token matcher cannot score, so it is excluded.)

\paragraph{Audio activation probe: no audio modality gap either, once the speech is intelligible.}
We replicate the vision activation probe's logit-lens on an omni model (Qwen2.5-Omni Thinker), presenting each novel fact as text vs as \emph{synthesized speech}, across three TTS engines of increasing quality (Table~\ref{tab:e7}). The result is a methodological caution turned finding: with low-quality TTS (espeak, only $71\%$ of clips intelligible to the model's own ASR) audio \emph{appears} to have a modality gap, but this tracks \emph{intelligibility}, not modality---with two independent clean neural voices ($88\%$ and $96\%$ intelligible) the gap vanishes and is TOST-equivalent within $\pm 1$ layer. Garbled TTS manufactures a spurious gap, so audio-modality claims must control for TTS intelligibility.

\begin{table}[h]
\small\centering
\caption{Audio activation probe readout (Qwen2.5-Omni Thinker, 24 novel facts). ``Intellig.'': fraction of clips the model's own ASR transcribes correctly. Depth-gap = audio$-$text readout depth (layers). TOST smallest effect size of interest (SESOI) $\pm 1$ layer.}
\label{tab:e7}
\setlength{\tabcolsep}{4pt}
\begin{tabular}{lccccc}
\toprule
TTS & intellig. & text acc & audio acc & depth-gap 90\% CI & TOST-equivalent \\
\midrule
espeak (robotic)        & 0.71 & 0.96 & 0.83 & $[0.64, 1.78]$ & no (spurious gap) \\
\texttt{mms-tts-eng} (clean)   & 0.88 & 0.96 & 1.00 & $[0.18, 0.57]$ & \textbf{yes} \\
\texttt{fish-speech-1.5} (natural) & 0.96 & 0.96 & 1.00 & $[0.21, 0.54]$ & \textbf{yes} \\
\bottomrule
\end{tabular}
\end{table}

\paragraph{Frontier retrieval probe (behavioral).}
The activation probes establish the modality null only on small open models (the lens needs activations). The frontier retrieval probe closes the gap behaviorally on the exact headline models: pack $K\in\{1,8,16,24,32,48\}$ novel nonsense-entity facts (distinct two-digit values, so prior knowledge gives no shortcut) as text or as an image of that text, and ask a single-number question about one target entity ($T{=}20$ seeded trials/cell). Table~\ref{tab:e8} gives the counts behind Figure~\ref{fig:substrate}b: on both models, legible-image accuracy equals text accuracy exactly at every $K$ up to 48, while the degraded positive control (small font, downscaled and blurred to low-DPI-scan quality) collapses and is decisively non-equivalent---the probe is sensitive.

\begin{table}[h]
\small\centering
\caption{Frontier retrieval probe: behavioral cross-modal test ($T{=}20$ seeded trials per $K$, single-number retrieval). ``legible img'': image of the same text at $34$px. ``degraded img'': small-font, downscaled+blurred low-DPI-scan control. Gap = image$-$text accuracy; TOST SESOI $\pm 0.1$.}
\label{tab:e8}
\setlength{\tabcolsep}{4pt}
\begin{tabular}{llcccc}
\toprule
Model & cond & acc @ $K{=}1$ & acc @ $K{=}48$ & pooled gap (95\% CI) & TOST-equiv \\
\midrule
\multirow{3}{*}{Gemini 3.1 Pro} & text          & 1.00 & 1.00 & --- & --- \\
                                & legible img   & 1.00 & 1.00 & $+0.000$ $[0.00,0.00]$ & \textbf{yes} \\
                                & degraded img  & 0.15 & 0.25 & $-0.733$ $[-0.81,-0.65]$ & no (control) \\
\midrule
\multirow{3}{*}{Claude Opus 4.7} & text         & 1.00 & 1.00 & --- & --- \\
                                 & legible img  & 1.00 & 1.00 & $+0.000$ $[0.00,0.00]$ & \textbf{yes} \\
                                 & degraded img & 0.15 & 0.25 & $-0.850$ $[-0.91,-0.78]$ & no (control) \\
\bottomrule
\end{tabular}
\end{table}

\paragraph{Frontier reasoning probe (multi-hop).}
The frontier reasoning probe forces composition: the model must pull $M{=}6$ specified facts out of a $K$-fact packed set and \emph{sum} them---the total is never a visible number---as text vs.\ legible image vs.\ the frontier retrieval probe's degraded control ($T{=}20$/cell over $K\in\{8,16,24,32\}$, Table~\ref{tab:e9}). On Gemini, text and legible image are both at ceiling and TOST-equivalent (gap $-0.013$). On Claude without extended thinking---the one sub-ceiling regime we could produce (text accuracy $0.825$; we could not drive clean-text Gemini below ceiling at any difficulty tried)---the legible image is reliably \emph{better} (image $1.00$, gap $+0.175$, 95\% CI $[0.10,0.26]$); the $14/80$ text errors are near-miss arithmetic slips that the image condition gets right. We read the Claude effect as a within-prompt echo of this paper's thesis---reading from an image nudges the model to externalize the numbers before summing---but report only the direction.

\begin{table}[h]
\small\centering
\caption{Frontier reasoning probe: retrieve $M{=}6$ packed facts and sum them (answer never directly visible), $T{=}20$/cell over $K\in\{8,16,24,32\}$. Gap = image$-$text accuracy; TOST SESOI $\pm 0.1$.}
\label{tab:e9}
\begin{tabular}{lcccc}
\toprule
Model & text acc & legible-img acc (gap) & TOST-equiv & degraded-img acc \\
\midrule
Gemini 3.1 Pro  & 1.00  & 0.99 ($-0.013$) & \textbf{yes} & 0.01 \\
Claude Opus 4.7 & 0.825 & \textbf{1.00 ($+0.175$)} & no (image \emph{better}) & 0.00 \\
\bottomrule
\end{tabular}
\end{table}

\paragraph{DPI and reading-order controls: they did not separate the hypotheses.}
The DPI control sought to vary vision-token count via DPI on a fixed document to probe the attention-dilution hypothesis. But Gemini normalizes images internally, so $96$ and $200$ DPI yield near-identical prompt token counts ($112{,}007$ vs $111{,}855$ on Heat~Pipes); the manipulation does not move the independent variable, and the attention-dilution hypothesis is untestable this way on Gemini. The reading-order control fed native digital text (\texttt{pdftotext}) to a single text-only call as a clean modality-hypothesis test, but on two-column arXiv PDFs \texttt{pdftotext} scrambles reading order, so native-text coverage is confounded downward (Fairness~AI $0.34$, below even $C_0$). Neither supports the modality hypothesis, so we report both with their confounds rather than omit them; Table~\ref{tab:mech_cond} collects the per-condition hallucination and coverage for all eight cells (the data behind Figure~\ref{fig:e2_collapse}).

\begin{table}[h]
\small\centering
\caption{Mechanistic conditions on the 8-cell subset (hallucinations\,h / probe-coverage); the data behind Figure~\ref{fig:e2_collapse}. $C_0$: end-to-end (modality). $C_1$: two-pass decomposition (text). $C_{1+}$: synthesis pass given both transcript and modality (modality-added control). single-call: transcribe-then-review in one response (single-call test, ``--'' = no review emitted, budget exhausted). native-text: single text-only call from \texttt{pdftotext} (reading-order control, arXiv only). Single-seed.}
\label{tab:mech_cond}
\begin{tabular}{lccccc}
\toprule
Cell & $C_0$ (img/aud) & $C_1$ (text) & $C_{1+}$ both & single-call & native-text \\
\midrule
MIT~6.034 (audio)        & 17h/0.77 & 4h/0.98 & --        & -- (budget) & -- \\
3B1B~Attention (audio)   & 11h/0.88 & 0h/0.96 & 5h/1.00   & 1h/0.80     & -- \\
Wagging~Tail         & 3h/0.80  & 1h/0.78 & 2h/0.78   & 5h/0.63     & -- \\
Natural~Vibration   & 4h/0.66  & 5h/0.49 & 9h/0.79   & -- (budget) & -- \\
Thermal             & 5h/0.68  & 2h/0.84 & 1h/0.68   & 3h/0.40     & -- \\
Heat~Pipes         & 14h/0.72 & 11h/0.78& 10h/0.66  & -- (budget) & -- \\
arXiv~Fairness      & 14h/0.50 & 15h/0.40& 11h/0.38  & 11h/0.24    & 13h/0.34 \\
arXiv~Perovskite    & 2h/0.90  & 3h/0.74 & 5h/0.84   & 5h/0.80     & 4h/0.80 \\
\bottomrule
\end{tabular}
\end{table}

\paragraph{Thinking-budget control: more reasoning does not substitute for a second pass.}
The single-call table above shows that forcing transcription into one response does not recover coverage. A stronger form of the ``it's just more budget'' objection is that the two-pass win is only about \emph{total} budget, so simply letting the one-pass model \emph{think} longer should close the gap. It does not. Holding the one-pass configuration fixed (same raw modality, temperature, and $65$k output cap), we sweep only Gemini's thinking budget across its full supported range---default (uncapped), then $128 \to 4{,}096 \to 32{,}768$ tokens, a $256\times$ span including the ceiling---on the eight-cell subset (Table~\ref{tab:e10}). For the paper cells we substitute a modality-neutral review prompt: the audio-worded headline prompt makes the model \emph{refuse} page images at the smallest budgets (an artifact unrelated to generation load, giving spurious zero-coverage outputs), which the neutral prompt removes. The result is the same everywhere. Coverage rises from the minimum budget up to the model's default level and then plateaus: maximum-budget coverage equals default-budget coverage within noise (mean difference $+0.005$ across the eight cells) and never approaches $C_1$. On the three cells where two-pass genuinely helps, the maximum reasoning budget closes at most a few points of an 8--21-point gap (MIT $0.77\!\to\!0.81$ vs.\ $C_1\,0.98$; Thermal $0.68\!\to\!0.72$ vs.\ $0.84$; 3B1B $0.88\!\to\!0.88$ vs.\ $0.96$). Capturing the model's reasoning traces (\texttt{include\_thoughts}) explains why: at every budget the reasoning is a \emph{plan} for the review---the model organizes the source into sections and decides what to emphasize---and never contains a verbatim transcript of the source. Extra thinking buys a better-organized but still satisficing review; it does not perform the externalization the separate Pass~1 does. This is the control that separates generation-under-load from a trivial budget effect: recovery requires a second \emph{output}, not a longer first one.

\begin{table}[h]
\small\centering
\caption{Thinking-budget control: one-pass probe coverage with only the reasoning budget varied. \emph{def} is default (uncapped) thinking; the rest are the nominal \texttt{thinking\_budget} ($128/4096/32768$). Audio cells use the headline review prompt; paper cells use a modality-neutral prompt (the audio-worded prompt makes the model refuse page images at the smallest budgets). Two-pass $C_1$ shown for reference. Maximum-budget coverage tracks the default and never reaches $C_1$.}
\label{tab:e10}
\begin{tabular}{llccccc}
\toprule
Cell & mod. & def & @128 & @4k & @32k & $C_1$ \\
\midrule
3B1B Attention    & audio & 0.88 & 0.80 & 0.80 & 0.88 & 0.96 \\
MIT 6.034         & audio & 0.77 & 0.46 & 0.83 & 0.81 & 0.98 \\
\midrule
Wagging Tail      & paper & 0.82 & 0.67 & 0.80 & 0.82 & 0.78 \\
Natural Vibration & paper & 0.79 & 0.68 & 0.68 & 0.77 & 0.49 \\
Thermal Analysis  & paper & 0.68 & 0.58 & 0.68 & 0.72 & 0.84 \\
Heat Pipes        & paper & 0.80 & 0.70 & 0.78 & 0.78 & 0.78 \\
arXiv Fairness    & paper & 0.46 & 0.24 & 0.34 & 0.46 & 0.40 \\
arXiv Perovskite  & paper & 0.86 & 0.70 & 0.88 & 0.86 & 0.74 \\
\bottomrule
\end{tabular}
\end{table}

\section{Vision activation probe: model ladder and parametric load sweep}
\label{appdx:vision_ladder}

Table~\ref{tab:vision_ladder_appdx} reports the vision activation probe's logit-lens readout for both open VLMs across the load sweep $K\in\{1,4,8,16,24\}$ (facts packed into one stimulus, one queried). Answer accuracy is $1.00$ for text and image at every $K$ on both models, and the text$-$image readout-depth gap stays within $\pm 1$ layer throughout (TOST-equivalent within a $\pm 1$-layer SESOI at every row). The Qwen2.5-VL-32B variant was excluded: it verbalizes the answer in a form our single-token matcher cannot capture (a spurious $\text{acc}=0$ artifact), so its readout depths are unreliable, and the 3B and 7B models carry the claim.

\begin{table}[h]
\centering
\small
\caption{Vision activation probe: cross-modal logit-lens across two open VLMs and a $K$-fact load sweep. \emph{text/img depth} = first layer the answer token is top-1. \emph{gap CI90} = 90\% CI on (img$-$text) depth in layers. \emph{equiv} = TOST-equivalent within $\pm 1$ layer. Accuracy is $1.00$ for both modalities in every row.}
\label{tab:vision_ladder_appdx}
\begin{tabular}{lrrrrrc}
\toprule
Model & $K$ & text acc & img acc & text depth & img depth & gap CI90 (equiv) \\
\midrule
Qwen2.5-VL-3B & 1  & 1.00 & 1.00 & 32.00 & 32.00 & $[0.00,0.00]$ (yes) \\
Qwen2.5-VL-3B & 4  & 1.00 & 1.00 & 32.25 & 32.06 & $[-0.41,0.04]$ (yes) \\
Qwen2.5-VL-3B & 8  & 1.00 & 1.00 & 32.19 & 32.06 & $[-0.27,0.02]$ (yes) \\
Qwen2.5-VL-3B & 16 & 1.00 & 1.00 & 32.75 & 32.06 & $[-0.88,-0.49]$ (yes) \\
Qwen2.5-VL-3B & 24 & 1.00 & 1.00 & 32.62 & 32.00 & $[-0.83,-0.42]$ (yes) \\
\midrule
Qwen2.5-VL-7B & 1  & 1.00 & 1.00 & 24.31 & 24.44 & $[-0.02,0.27]$ (yes) \\
Qwen2.5-VL-7B & 4  & 1.00 & 1.00 & 24.25 & 24.12 & $[-0.27,0.02]$ (yes) \\
Qwen2.5-VL-7B & 8  & 1.00 & 1.00 & 24.00 & 24.00 & $[-0.15,0.15]$ (yes) \\
Qwen2.5-VL-7B & 16 & 1.00 & 1.00 & 24.06 & 24.12 & $[-0.04,0.17]$ (yes) \\
Qwen2.5-VL-7B & 24 & 1.00 & 1.00 & 24.06 & 24.12 & $[-0.04,0.17]$ (yes) \\
\bottomrule
\end{tabular}
\end{table}

\section{Protocol details}
\label{appdx:protocol}

\paragraph{Reference quality.} Gemini~3.1~Pro is used for OCR/ASR with chunking. Single-call OCR on a 104-page paper produced a degenerate TOC dot-leader loop (113KB output, 56k ``words'' but only 264 real-word tokens, 6 page markers out of 104). Chunking (1 page per call, max 8192 output tokens) eliminates this failure mode. For audio we chunk into 30-min segments with 5-second overlap. We discovered a silent server-side truncation on one audio chunk (chunk~1 of IETF SNAC produced 1981 words instead of the expected $\approx 4368$, ending mid-sentence with \texttt{``Uh, well, see, yeah, that'\,''}). The protocol grew an automatic chunk-level word-count parity check after this incident.

\paragraph{Statistics.} We treat $|\Delta h|\le 1$ on cells with $h<10$, and $|\Delta\mathrm{cov}|\le 0.04$ ($\approx 2$ probes on a 50-probe cell), as within noise (presentation conventions calibrated against inter-judge direction flips, \S\ref{sec:robust}). For aggregate effects we report Wilson 95\% CIs on coverage rates, percentile-bootstrap 95\% CIs ($n_{\rm boot}{=}10000$, seeded) on cross-cell means and Pearson $r$, and sign tests against $\Delta{=}0$. The judge runs with a 24k-token completion budget.

\paragraph{Sources.} Audio: Karpathy ``State of GPT'' (42 min, 2023), MIT~6.034 L1 Patrick Winston (47 min, 2010), Sapolsky Behavioral~Genetics~1 Stanford (85 min, post-2026 reupload), Stanford Decarbonized~Power~Sector Zach~Ming (72 min, 2025), Doudna CRISPR Basics (48 min, 2017), 3Blue1Brown ``Attention in Transformers'' (26 min, 2024), Veritasium ``Strange Math'' (32 min, 2025), NeurIPS Black-in-AI panel (31 min, 2021), Harvard Law School Ames Moot Court (82 min, 2025), IETF SNAC WG interim 2026-04-16 (60 min, $\approx 5$ speakers, recorded post the most likely training-data cutoff). Karpathy yields both a review and a meeting-minutes task. Papers: NASA NTRS IDs 19720009221 (Wagging Tail~\cite{nasa_waggingtail}), 19680000724 (Env Test~\cite{nasa_envtest}), 19690008405 (Dynamic Response~\cite{nasa_dynresp}), 19690013408 (Natural Vibration~\cite{nasa_naturalvib}), 19700023812 (Thermal Analysis~\cite{nasa_thermal}), 19700025120 (Heat Pipes~\cite{nasa_heatpipes}), and SP-5100 (Shock and Vibration Technology~\cite{nasa_sp5100}). The arXiv papers are arXiv:2605.09852 (Fairness~AI~\cite{arxiv_fairness}), arXiv:2605.13991 (Perovskite Tandem PV~\cite{arxiv_perovskite}), and arXiv:2605.11379 (Megagauss Physics~\cite{arxiv_megagauss}). NASA papers are rendered as page images at 200 DPI, and arXiv papers at 200 DPI full color.

\paragraph{Models and snapshots.} The models are Gemini 3.1 Pro Preview (\texttt{gemini-3.1-pro-preview}), Claude Opus 4.7 (\texttt{claude-opus-4-7}), and Mimo v2-omni (\texttt{xiaomi/mimo-v2-omni} via OpenRouter). The judge is GPT-5.4 (\texttt{reasoning\_effort=high}, completion budget 24k tokens), routed via OpenRouter due to scope restrictions on the local OpenAI key. Cross-vendor extensions reuse the Gemini reference and probes, isolating the question to whether the advantage depends on the $C_0$/$C_1$ model: Claude Opus 4.7 (complete $C_0$-vs-$C_1$ on the six papers where its multimodal $C_0$ fits; $C_1$-only on the longer papers) and Mimo v2-omni on every cell where its $C_0$ does not return empty.

\paragraph{Mimo audio constraint.} OpenRouter's Mimo provider enforces a 10MB base64 audio input limit. Karpathy required re-encoding from 32kbps to 16kbps mono mp3 to fit, and Sapolsky at 85 min required 12kbps. The Mimo audio Karpathy result ($C_0$ 4h/0.569, $C_1$ 4h/0.588 across 51 probes, 1509 / 1268 review words) is within noise across the $C_0\to C_1$ transition, so we treat it as a wash. \emph{gpt-audio} (via OpenRouter) accepted audio input but refused text-only Pass-2 with HTTP 400 ``This model requires that either input content or output modality contain audio''. The same-weights cascade is not realizable on this model, and $C_0$ alone scored 10h/0.765 on Karpathy.

\paragraph{Pareto terminology.} We call a cascade condition \emph{strictly Pareto-positive} relative to $C_0$ when it has both fewer hallucinations and higher coverage by amounts exceeding the noise floor. \emph{Pareto-positive within noise} means strictly better on one axis and no worse than noise on the other. \emph{Pareto-dominant} on a set means no other condition has lower hallucination at equal or higher coverage. These are presentation labels, not statistical claims.

\section{All 21 cells: raw counts and Wilson 95\% CIs}
\label{appdx:cell_table}

\begin{figure}[h]
\centering
\includegraphics[width=0.62\linewidth]{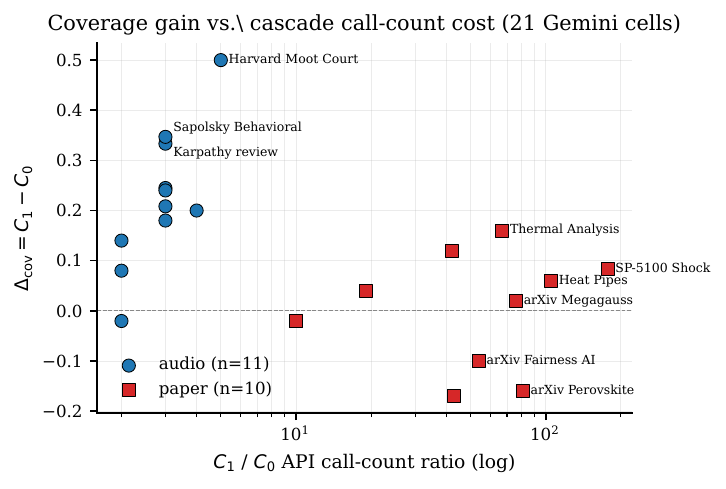}
\caption{\textbf{Coverage gain versus API call-count cost.} Each point is a Gemini cell. Paper $C_1/C_0$ call ratios are $N{+}1$ (one OCR call per page plus Pass~2), versus roughly $2$--$3\times$ for chunked audio ASR. Gains are generally largest for cheap audio and smaller for costly papers, but baseline headroom explains exceptions such as SP-5100.}
\label{fig:cost}
\end{figure}

Figure~\ref{fig:cost} plots each Gemini cell's coverage gain against its API call-count cost, and Table~\ref{tab:all_cells} lists the raw per-cell counts with Wilson 95\% CIs.

\begin{table}[h]
\centering
\scriptsize
\caption{All 21 Gemini base cells (data for Figures~\ref{fig:headline} and \ref{fig:headline_cells}). Halluc.\ is unsupported-claim count; cov.\ is covered probes $k/n$ (Wilson 95\% CI). $\dagger$ marks cells without an above-noise improvement on at least one axis. \emph{cite/1k} counts numeric citation surface forms per 1k source words and therefore undercounts author--year citations.}
\label{tab:all_cells}
\resizebox{\textwidth}{!}{%
\begin{tabular}{lrrrrrrr}
\toprule
Cell & len & $C_0$ halluc & $C_0$ cov [95\% CI] & $C_1$ halluc & $C_1$ cov [95\% CI] & $\Delta_{\rm cov}$ & cite/1k \\
\midrule
\multicolumn{8}{l}{\emph{Audio (11 cells)}} \\
Karpathy review        & 42m & 13 & 0.627 [0.490, 0.747] & 3  & 0.961 [0.868, 0.989] & $+0.333$ & 0.00 \\
Karpathy mtg-min       & 42m & 4  & 0.160 [0.083, 0.285] & 2  & 0.340 [0.224, 0.478] & $+0.180$ & 0.00 \\
SNAC mtg-min           & 60m & 14 & 0.286 [0.178, 0.424] & 11 & 0.531 [0.394, 0.663] & $+0.245$ & 0.00 \\
3B1B Attention         & 26m & 11 & 0.880 [0.762, 0.944] & 0  & 0.960 [0.864, 0.989] & $+0.080$ & 0.00 \\
Veritasium Math        & 32m & 14 & 0.820 [0.692, 0.903] & 6  & 0.960 [0.864, 0.989] & $+0.140$ & 0.00 \\
MIT 6.034 Winston      & 47m & 17 & 0.771 [0.633, 0.866] & 4  & 0.979 [0.890, 0.996] & $+0.208$ & 0.00 \\
NeurIPS BinAI panel    & 31m & 6  & 0.918 [0.803, 0.969] & 4  & 0.898 [0.778, 0.957] & $-0.020^{\dagger}$ & 0.00 \\
Doudna CRISPR          & 48m & 6  & 0.680 [0.542, 0.792] & 3  & 0.920 [0.812, 0.968] & $+0.240$ & 0.00 \\
Stanford Decarb        & 72m & 11 & 0.700 [0.562, 0.809] & 6  & 0.900 [0.786, 0.957] & $+0.200$ & 0.00 \\
Harvard Moot Court     & 82m & 11 & 0.360 [0.241, 0.499] & 4  & 0.860 [0.736, 0.932] & $+0.500$ & 0.00 \\
Sapolsky Behavioral    & 85m & 19 & 0.592 [0.450, 0.720] & 15 & 0.939 [0.834, 0.980] & $+0.347$ & 0.00 \\
\midrule
\multicolumn{8}{l}{\emph{Paper, NASA NTRS 1968--1972}} \\
Wagging Tail          &   9p & 3  & 0.796 [0.664, 0.885] & 1  & 0.776 [0.641, 0.870] & $-0.020^{\dagger}$ & 1.42 \\
Env Test              &  18p & 5  & 0.880 [0.762, 0.944] & 4  & 0.920 [0.812, 0.968] & $+0.040^{\dagger}$ & 0.00 \\
Dynamic Response      &  41p & 3  & 0.580 [0.442, 0.706] & 2  & 0.700 [0.562, 0.809] & $+0.120$ & 1.19 \\
Natural Vibration     &  42p & 4  & 0.660 [0.517, 0.778] & 5  & 0.489 [0.353, 0.628] & $-0.170^{\dagger}$ & 6.33 \\
Thermal Analysis      &  66p & 5  & 0.680 [0.542, 0.792] & 2  & 0.840 [0.715, 0.917] & $+0.160$ & 0.36 \\
Heat Pipes            & 104p & 14 & 0.720 [0.583, 0.825] & 11 [med 12, $\{11,12,17\}$] & 0.780 [0.648, 0.872] & $+0.060$ & 2.80 \\
SP-5100 Shock         & 176p & 47 [med 46.5] & 0.521 [0.383, 0.655] & 45 [med 45, $\{25,45,59\}$] & 0.604 [0.463, 0.730] & $+0.083$ & 4.00 \\
\midrule
\multicolumn{8}{l}{\emph{Paper, arXiv 2026 (post-cutoff)}} \\
arXiv Fairness AI     &  53p & 14 & 0.500 [0.366, 0.634] & 15 & 0.400 [0.276, 0.538] & $-0.100^{\dagger}$ & 0.03 \\
arXiv Megagauss       &  75p & 36 [$\{36,36\}$] & 0.660 [0.522, 0.776] & 32 [med 30, $\{28,32\}$] & 0.680 [0.542, 0.792] & $+0.020^{\dagger}$ & 9.46 \\
arXiv Perovskite      &  80p & 2  & 0.900 [0.786, 0.957] & 3  & 0.740 [0.604, 0.841] & $-0.160^{\dagger}$ & 0.15 \\
\bottomrule
\end{tabular}%
}
\end{table}

\paragraph{A held-out predictor (LOOCV, $n{=}21$).}
A leave-one-out logistic-regression predictor on four cell-level features ($C_0$ probe coverage, $\log_{10}$ source words, $C_0$ unsupported-claim count, and a paper/audio dummy) reaches LOOCV AUC $= 0.65$ on $\text{sign}(\Delta_{\rm cov}>0)$. All four coefficient signs match the qualitative picture (higher baseline reduces cascade benefit, longer source reduces benefit, more $C_0$ hallucinations \emph{increases} benefit, paper modality harder than audio). AUC $0.65$ is meaningful but well below deployment-ready, so we report it as an honest $n{=}21$ baseline. Adding a citation-density feature drops AUC to $0.61$: the residual Mode-B structure is not captured by a single density feature at this sample size (and the feature itself undercounts author--year citation styles; Table~\ref{tab:all_cells}).

\section{Cross-vendor details}
\label{appdx:xvendor}

\begin{figure}[h]
\centering
\includegraphics[width=0.55\linewidth]{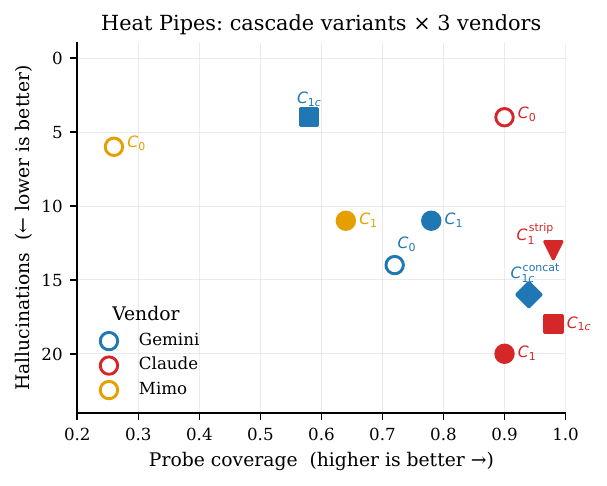}
\caption{Heat Pipes Pareto frontier across vendors and cascade variants. The Gemini cascade is small-Pareto-positive; Claude $C_1$ regresses on faithfulness, leaving $C_0$ on the Pareto frontier (the Mode-B signature); Mimo trades faithfulness for coverage (Modes A+B compounded).}
\label{fig:xvendor_appdx}
\end{figure}

Figure~\ref{fig:xvendor_appdx} shows the Heat~Pipes Pareto frontier across vendors and cascade variants, and Table~\ref{tab:claude_within} gives the within-Claude $C_0$--$C_1$ comparison on the six NASA papers that fit its multimodal payload.

\paragraph{SP-5100 replication.} NASA SP-5100 ``Shock and Vibration Technology'' (1972, 47k pdftotext words, 4.00 citation surface forms / 1k words) was chosen as a second long citation-heavy NASA survey for Mode~B replication. Gemini $C_0$ median $46.5$h/0.52 cov $\to C_1$ median $45$h/0.60 cov (range $\{25,45,59\}$ across three judge re-runs of the byte-identical review, i.e.\ judge variance, not Pass-2 variance). Claude $C_1$ median $85$h/0.81 cov (range $\{47,85,85,97\}$ across four judge re-runs, where the originally reported $47$h was a low-end outlier). Mimo $C_1\,60$h/0.23 cov. Claude $C_0$ not realizable: the page-images exceed Anthropic's per-request payload limit even with the 1M-context beta. Spot-checking Claude $C_1$ shows the same bibliographic-expansion signature as Heat~Pipes.

\begin{table}[h]
\centering
\small
\caption{\textbf{Within-Claude $C_0$--$C_1$ comparison} on the six NASA papers that fit its multimodal payload (data for Figure~\ref{fig:xmodel}b). Halluc.\ is the GPT-5.4 unsupported-claim count; cov.\ uses the Gemini reference/probes. Cells are sorted by length.}
\label{tab:claude_within}
\begin{tabular}{lrrrrr}
\toprule
Cell & len & $C_0$ h/cov & $C_1$ h/cov & $\Delta_h$ & $\Delta_{\rm cov}$ \\
\midrule
Wagging Tail      &   9p & 5 / 0.959  & 2 / 0.980  & $-3$  & $+0.021$ \\
Env Test          &  18p & 3 / 1.000  & 2 / 1.000  & $-1$  & $\pm0.000$ \\
Dynamic Response  &  41p & 2 / 1.000  & 1 / 0.960  & $-1$  & $-0.040$ \\
Natural Vibration &  42p & 5 / 0.915  & 4 / 0.851  & $-1$  & $-0.064$ \\
Thermal Analysis  &  66p & 11 / 0.920 & \textbf{4 / 1.000} & $-7$ & $+0.080$ \\
Heat Pipes        & 104p & \textbf{4} / 0.900 & 20 / 0.900 & $+16$ & $\pm0.000$ \\
\midrule
\multicolumn{4}{l}{\emph{direction (noise floor $|\Delta_h|{\ge}1$, $|\Delta_{\rm cov}|{\ge}0.04$)}} & 5/6 $\downarrow$ & --- \\
\bottomrule
\end{tabular}
\end{table}

\paragraph{Modern arXiv reviews.} Gemini Fairness~AI: $14$h/0.50 $\to$ $15$h/0.40 ($-10$pp coverage, Mode~A). Claude $C_1\,23$h/0.70, and Mimo $C_1\,18$h/0.32. Gemini Megagauss: $C_0$ median $36$h/0.66 $\to C_1$ median $30$h/0.68 (high baseline-halluc physics review). Claude $C_1\,17$h/0.92, and Mimo $C_1\,19$h/0.26. Gemini Perovskite: $2$h/0.90 $\to$ $3$h/0.74 at $n{=}1$ ($-6$pp under multi-seed, near-noise-floor). Claude $C_1\,0$h/0.90, and Mimo $C_1\,2$h/0.90. The arXiv pattern (Mode~A on $2/3$) is opposite-direction from the NASA pattern despite comparable page counts. We attribute this to (a)~arXiv full-color 200~DPI rendering carrying more visual signal per page, and (b)~higher typesetting density forcing more aggressive $C_1$ compression at the same output budget.

\section{Mixed-model audio cascade: decomposition unlocks text-only models}
\label{appdx:mixed}

The same-weights audio cascade is Gemini-only by API constraint (\S\ref{sec:discussion}). In production one would not use one model for everything; one would use the best model per stage. This probe does exactly that, and is therefore \emph{off-protocol}: it mixes vendors across stages, so it speaks to the generality of \emph{decomposition}, not the same-weights claim. We take an \emph{independent}, non-Google Pass-1 transcript from OpenAI \texttt{gpt-audio} (chunked, full quality---distinct from the Gemini reference used for judging), then run a text-only synthesis Pass-2 with three frontier text models, two of which (Claude Opus 4.7, GPT-5.4) cannot ingest audio at all on their own. Reviews are scored on the shared Pro reference and probes with the GPT-5.4 judge, plus a Claude-judge cross-check on the GPT-5.4-authored review to rule out self-grading.

The result is unambiguous (Table~\ref{tab:mixed}; Figure~\ref{fig:xmodel}d): on all three cells, every mixed pipeline beats Gemini's \emph{end-to-end} $C_0$ on both axes---coverage rises to $0.85$--$1.00$ (vs $C_0$ $0.63$--$0.88$) and hallucinations fall to $2$--$9$ (vs $11$--$17$). Decomposition thus \emph{unlocks} text-only frontier models for audio review and lets them outperform the native audio model's end-to-end pass, using an off-the-shelf transcript. The GPT-5.4 self-grading concern is unfounded: the Claude judge scores the GPT-5.4-authored reviews \emph{lower} on hallucinations (1/3/3 vs the GPT-5.4 judge's 3/6/9), consistent with the $0.71\times$ inter-judge ratio (\S\ref{sec:robust}). Where the mixed pipeline trails the Gemini same-weights $C_1$ (most clearly on MIT: coverage $0.85$--$0.90$ vs $0.98$ across all three Pass-2 models), the deficit tracks \texttt{gpt-audio}'s lossier transcript, i.e.\ the ceiling is set by Pass-1 transcript quality, exactly as the perceive--externalize--synthesize account predicts.

\begin{table}[h]
\centering
\small
\caption{Mixed-model audio cascade (off-protocol): independent \texttt{gpt-audio} Pass-1 transcript $\to$ text-only Pass-2 by each model, vs Gemini~3.1~Pro end-to-end $C_0$. Hallucinations\,h / probe-coverage, GPT-5.4 judge; the parenthetical on GPT-5.4 is the Claude-judge cross-check. Claude and GPT-5.4 take no audio input, so their only route to an audio review is via the cascade.}
\label{tab:mixed}
\setlength{\tabcolsep}{3pt}
\begin{tabular}{lcccc}
\toprule
Cell & Gemini $C_0$ (e2e) & \texttt{gpt-audio}$\to$Claude & \texttt{gpt-audio}$\to$GPT-5.4 & \texttt{gpt-audio}$\to$Gemini \\
\midrule
3B1B Attention & 11h / 0.88 & \textbf{2h / 1.00} & 3h / 1.00 \,(1h / 1.00) & 5h / 1.00 \\
Karpathy       & 13h / 0.63 & 6h / 0.98 & 6h / 1.00 \,(3h / 1.00) & 5h / 0.90 \\
MIT 6.034      & 17h / 0.77 & 3h / 0.90 & 9h / 0.90 \,(3h / 0.92) & 5h / 0.85 \\
\bottomrule
\end{tabular}
\end{table}

\section{Cross-model audio: Gemini 2.5 Flash (clean, no bitrate confound)}
\label{appdx:flash_audio}

Table~\ref{tab:flash_audio} reports a full-quality $C_0$-vs-$C_1$ comparison with Gemini~2.5~Flash as the subject model (same reused Pro reference and probes, same GPT-5.4 judge), over five representative audio cells; Figure~\ref{fig:xmodel}c plots it against the Pro arm. Unlike the Mimo audio arm, the audio is the full 32~kbps source (no OpenRouter 10MB cap), so there is no audio-quality confound. 2.5~Flash is a strong end-to-end audio reader: its $C_0$ coverage exceeds Gemini~3.1~Pro's on all five cells (only on Harvard does it pay for that coverage with more hallucinations, 19 vs Pro's 11), leaving little headroom. The cascade direction then follows the same headroom pattern---a clear win on the two headroom cells (MIT, Harvard), flat on the saturated NeurIPS panel, and a faithfulness regression on Karpathy (where $C_0$ is already $0.92$/3h). Net: faithfulness improves on 3/5 and coverage on 1/5, a smaller and headroom-gated effect rather than the uniform audio win seen on Pro. Gemini~3~Flash was attempted as a second subject but its verbatim Pass-1 transcription returns empty under the model's recitation filter on copyrighted audio, so the cascade is not realizable on it.

\begin{table}[h]
\centering
\small
\caption{Gemini 2.5 Flash audio $C_0$-vs-$C_1$ (five cells, full-quality audio, GPT-5.4 judge on the reused Pro reference/probes). Sorted by length.}
\label{tab:flash_audio}
\begin{tabular}{lrrrrr}
\toprule
Cell & min & $C_0$ h/cov & $C_1$ h/cov & $\Delta_h$ & $\Delta_{\rm cov}$ \\
\midrule
3B1B Attention & 26 & 5 / 0.900  & 1 / 0.920  & $-4$ & $+0.020$ \\
NeurIPS panel  & 31 & 5 / 0.980  & 6 / 0.959  & $+1$ & $-0.020$ \\
Karpathy       & 42 & 3 / 0.922  & 9 / 0.902  & $+6$ & $-0.020$ \\
MIT 6.034      & 47 & 7 / 0.833  & \textbf{4 / 0.958} & $-3$ & $+0.125$ \\
Harvard Moot   & 82 & 19 / 0.600 & \textbf{12 / 0.620} & $-7$ & $+0.020$ \\
\bottomrule
\end{tabular}
\end{table}

\section{Iterative refinement ($C_{1\rm iter}$): full results}
\label{appdx:iter}

\begin{figure}[h]
\centering
\includegraphics[width=0.72\linewidth]{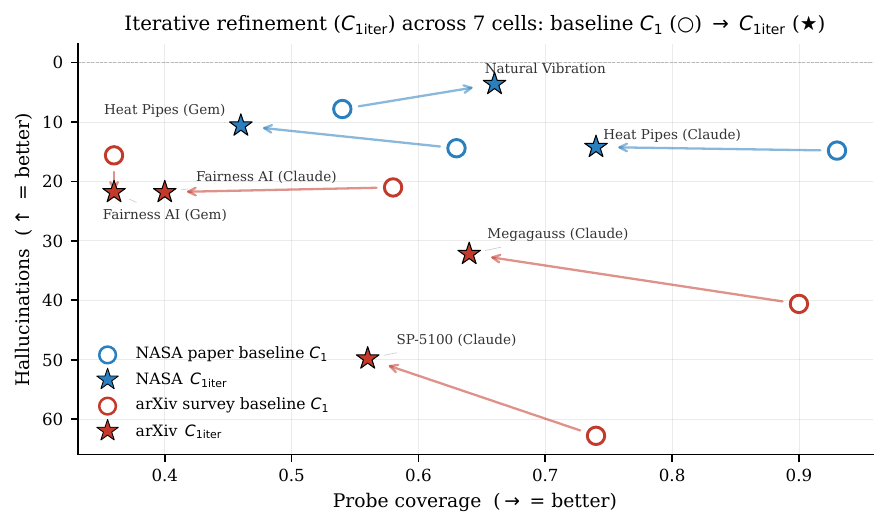}
\caption{\textbf{Quote-grounded iterative refinement ($C_{1\rm iter}$) across seven cells ($n{=}5$ means).} Stars show $C_{1\rm iter}$ relative to plain $C_1$ (open circles). Natural~Vibration is the only clean Pareto win; other cells trade coverage against faithfulness or regress, depending on whether Mode~B is the binding constraint.}
\label{fig:iter_pareto}
\end{figure}

The three-step pipeline and the high-level verdict are in \S\ref{sec:failure_modes}; Figure~\ref{fig:iter_pareto} plots the per-cell movement and Table~\ref{tab:iter_results} gives the $n{=}5$ numbers.

\begin{table}[h]
\centering
\small
\caption{Iterative refinement ($C_{1\rm iter}$) on seven cells at $n{=}5$, $T{=}0.7$. Hallucinations are mean$\pm$sd and coverage is mean; $\Delta$ is versus a plain $C_1$ baseline re-measured at $n{=}5$. ``Pareto'' improves both axes beyond noise; ``Trade'' exchanges them; ``Net-neg'' loses coverage without a faithfulness gain.}
\label{tab:iter_results}
\begin{tabular}{lrrrrl}
\toprule
Cell (vendor) & $C_1$ baseline & $C_{1\rm iter}$ & $\Delta_{\rm halluc}$ & $\Delta_{\rm cov}$ & Verdict \\
\midrule
Natural Vibration (Gemini) & $7.8{\pm}2.2$ / $0.54$  & $\mathbf{3.6{\pm}1.4}$ / $\mathbf{0.66}$ & $-4.2$  & $+0.13$ & \textbf{Pareto} \\
Heat Pipes (Gemini)        & $14.4{\pm}2.6$ / $0.63$ & $10.6{\pm}3.1$ / $0.46$ & $-3.8$  & $-0.17$ & trade   \\
Heat Pipes (Claude)        & $14.8{\pm}3.5$ / $0.93$ & $14.2{\pm}4.4$ / $0.74$ & $-0.6$  & $-0.20$ & net-neg \\
arXiv Fairness AI (Gemini) & $15.6{\pm}6.6$ / $0.36$ & $21.8{\pm}1.0$ / $0.36$ & $+6.2$  & $+0.01$ & net-neg \\
arXiv Fairness AI (Claude) & $21.0{\pm}3.9$ / $0.58$ & $21.8{\pm}3.2$ / $0.40$ & $+0.8$  & $-0.17$ & net-neg \\
SP-5100 (Claude)           & $62.8{\pm}23$ / $0.74$  & $49.8{\pm}13$ / $0.56$ & $-13.0$ & $-0.18$ & trade   \\
arXiv Megagauss (Claude)   & $40.6{\pm}4.6$ / $0.90$ & $32.2{\pm}8.5$ / $0.64$ & $-8.4$  & $-0.25$ & trade   \\
\bottomrule
\end{tabular}
\end{table}

$C_{1\rm iter}$ targets Mode~B (a verbatim-quote requirement rules out unsupported expansions) but does not relieve Mode~A: the rewrite step itself compresses, so coverage falls on every long cell, and the net verdict is the balance of recovered faithfulness against that compression cost. Two single-seed verdicts did not survive $n{=}5$: the Heat~Pipes~Claude Mode-B ``fix'' (now within noise) and an apparent ``hallucinations increase'' on SP-5100/Megagauss (both counts in fact \emph{fall} at $n{=}5$; the single-seed rises were high-variance GPT-5.4 judge draws on 40--90-claim lists).

\section{Engineering interventions: chunked Pass-2 and citation stripping}
\label{appdx:fixes}

Table~\ref{tab:hp_claude_variants} lists the Heat~Pipes Claude cascade variants at $n{=}1$.

\begin{table}[h]
\centering
\caption{Heat~Pipes Claude cascade variants ($n{=}1$).}
\label{tab:hp_claude_variants}
\small
\begin{tabular}{lrrr}
\toprule
Variant & Halluc. & Coverage & Words \\
\midrule
$C_0$ (image baseline) & \textbf{4}  & 0.900 & 3{,}308 \\
$C_1$ (single Pass-2) & 20 & 0.900 & 2{,}077 \\
$C_1^{\rm strip}$ (citations $\to$ \texttt{[REF]}) & \textbf{13} & \textbf{0.980} & 2{,}471 \\
$C_{1c}$ (chunked + merge) & 18 & 0.980 & 2{,}769 \\
$C_{1c}^{\rm concat}$ (chunked, no merge) & 20 & 0.940 & 2{,}777 \\
$C_{1c}^{\rm strip}$ (strip + chunked + merge) & 18 & 0.980 & 2{,}673 \\
$C_{1c}^{\rm strip+concat}$ (strip + chunked, no merge) & 16 & 0.980 & 2{,}766 \\
\bottomrule
\end{tabular}
\end{table}

\paragraph{Mode-B fix replication on 3 new long Claude cells.}
Table~\ref{tab:mode_b_strip_replication} reports the strip intervention across four Claude cells. On SP-5100, the published single-call snapshot $C_1\,47h \to C_1^{\rm strip}\,24h$ initially looked like a clean halving, but the unstripped $C_1$ review judged four times yields $\{47, 85, 85, 97\}$ (median $85$), so the $-23$h effect compares one strip call against a low-end outlier of the unstripped distribution. On Megagauss, strip has essentially no effect ($17\to 19$h, $0.92\to 0.94$ cov). On Fairness~AI only 1 surface form matched our regexes (its citations are author--year, which the numeric-form patterns do not match, so the intervention never touched the text it targets), and the ``stripped'' Pass-2 produced a substantially different review ($23\to 41$h, $0.70\to 0.62$ cov), revealing that Claude's text-only Pass-2 is itself non-deterministic. \textbf{Net Mode-B-fix status: zero replicated cells} (Heat~Pipes retracted at $n{=}5$, Megagauss null, Fairness~AI null-by-design, SP-5100 single-call-each).

\begin{table}[h]
\centering
\caption{Mode~B citation-strip intervention across four long Claude cells. ``surface forms'' = regex hits replaced with \texttt{[REF]}.}
\label{tab:mode_b_strip_replication}
\small
\begin{tabular}{lrrrrrr}
\toprule
Cell & len & surface forms & $C_1$ h & $C_1$ cov & $C_1^{\rm strip}$ h & $C_1^{\rm strip}$ cov \\
\midrule
Heat~Pipes (recall)        & 104p & 55  & 20 & 0.900 & \textbf{13} & \textbf{0.980} \\
SP-5100 (new)              & 176p & 193 & 47 & 0.812 & \textbf{24} & 0.708 \\
arXiv Megagauss (new)      &  75p & 298 & 17 & 0.920 & 19 & 0.940 \\
arXiv Fairness~AI (new)    &  53p & 1   & 23 & 0.700 & 41 & 0.620 \\
\bottomrule
\end{tabular}
\end{table}

\paragraph{Mode-A chunked Pass-2 replication and multi-seed verification.}
$n{=}1$ ($T{=}0$) results are consistent across four papers (Table~\ref{tab:mode_a_chunked_replication}): $C_{1c}^{\rm concat}$ (chunked, no merge) recovers coverage on long sources at a hallucination cost, while $C_{1c}$ (chunked + merge) over-compresses: on Perovskite and Megagauss the merge variant collapses to $0.10$ and $0.12$ coverage respectively. At $n{=}5$ multi-seed (Table~\ref{tab:chunked_multiseed_appdx}), the chunked-Pass-2 direction survives on Fairness~AI and Perovskite but \emph{not} on Megagauss, whose original $n{=}1$ ($0.74$cov) sits above the entire $n{=}5$ envelope $[0.64, 0.68]$ at $T{=}0.7$. We retract the original ``chunked Pass-2 lifts coverage on all three arXiv counter-cells'' claim.

\begin{table}[h]
\centering
\caption{Chunked-Pass-2 ($n{=}1$, $T{=}0$) across four Gemini long-paper cells.}
\label{tab:mode_a_chunked_replication}
\small
\begin{tabular}{lrrrrr}
\toprule
Cell & len & $C_1$ h/cov & $C_{1c}$ h/cov & $C_{1c}^{\rm concat}$ h/cov & $\Delta_{\rm cov}^{\rm concat}$ \\
\midrule
Heat~Pipes (recall)    & 104p & 11 / 0.78 & 4 / 0.58 & 16 / 0.94 & $+0.16$ \\
arXiv Perovskite (new) &  80p & 3  / 0.74 & 1 / 0.10 & 13 / 0.82 & $+0.08$ \\
arXiv Megagauss (new)  &  75p & 32 / 0.68 & 0 / 0.12 & 36 / 0.74 & $+0.06$ \\
arXiv Fairness~AI (new)&  53p & 15 / 0.40 & 5 / 0.32 & 37 / 0.52 & $+0.12$ \\
\bottomrule
\end{tabular}
\end{table}

\begin{table}[h]
\centering
\small
\caption{Multi-seed verification of $C_{1c}^{\rm concat}$ on the three arXiv cells ($n{=}5$ seeds, Gemini~3.1~Pro Pass-2, $T{=}0.7$).}
\label{tab:chunked_multiseed_appdx}
\resizebox{\textwidth}{!}{%
\begin{tabular}{lrrrrrl}
\toprule
Cell & $C_1$ baseline & orig $n{=}1$ & $n{=}5$ halluc & $n{=}5$ cov & cov range & dir.\ survives \\
\midrule
arXiv Fairness AI & 15h / 0.40 & 37h / 0.52 & $31.6\pm1.9$ & $0.496\pm0.055$ & $[0.440, 0.580]$ & yes \\
arXiv Megagauss   & 32h / 0.68 & 36h / 0.74 & $42.2\pm4.3$ & $0.664\pm0.017$ & $[0.640, 0.680]$ & \textbf{NO} \\
arXiv Perovskite  & 3h / 0.74  & 13h / 0.82 & $16.6\pm2.5$ & $0.832\pm0.033$ & $[0.800, 0.880]$ & yes \\
\bottomrule
\end{tabular}%
}
\end{table}

\paragraph{Citation stripping on Gemini Natural Vibration ($n{=}5$ verified).}
At $n{=}1$ ($T{=}0$), strip moves halluc $5\to 12$ but coverage $0.49\to 0.81$ ($+32$pp). At $n{=}5$ ($T{=}0.7$, seeds $1$--$5$): halluc $\{15,11,9,8,8\}$ with mean $10.2$, and coverage $\{0.660, 0.723, 0.702, 0.511, 0.766\}$ with mean $0.672$. The strip-vs-$C_1$ direction (cov gain at halluc cost) survives all 5 seeds. The magnitude on coverage is $+18$pp (mean) versus the original $+32$pp single-seed reading.

\section{Robustness of the headline result: the four checks in full}
\label{appdx:robustness}

The headline coverage/faithfulness effect (\S\ref{sec:robust}) is tested against four measurement threats. This appendix collects the four checks; the detailed per-cell tables follow in Appendices~\ref{appdx:multiseed}, \ref{appdx:inter_judge}, and~\ref{appdx:probe_circ}.

\begin{figure}[h]
\centering
\includegraphics[width=0.7\linewidth]{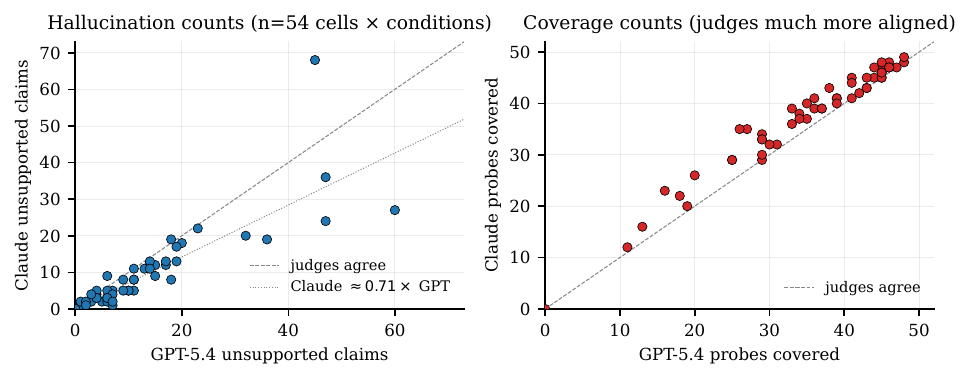}
\caption{\textbf{Threat 1 --- inter-judge replication} on all 21 cells ($+$1 cross-vendor cell). A Claude Opus~4.7 re-judge agrees with GPT-5.4 on most coverage and hallucination directions.}
\label{fig:rob_a}
\end{figure}

\emph{Threat 1: judge idiosyncrasy.} We re-judge every cell with a second frontier judge (Claude Opus~4.7) in place of GPT-5.4. The two judges agree on the direction of the coverage and hallucination changes on most cells (Figure~\ref{fig:rob_a}); the overall inter-judge hallucination ratio is $0.71\times$, i.e.\ Claude counts somewhat fewer unsupported claims, but the sign of the effect is preserved.

\begin{figure}[h]
\centering
\includegraphics[width=0.7\linewidth]{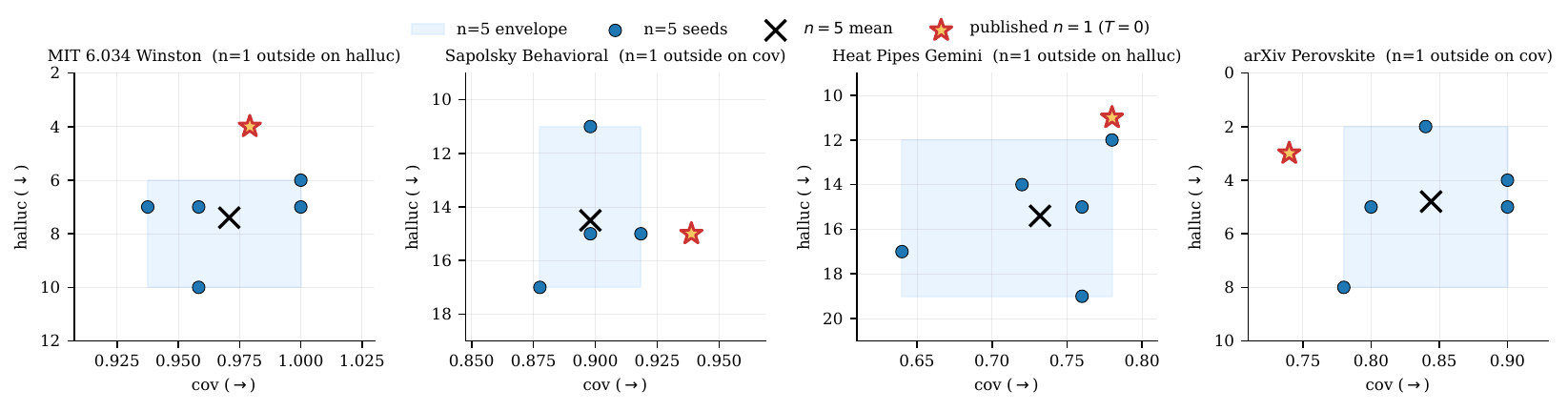}
\caption{\textbf{Threat 2 --- $n{=}5$ Pass-2 multi-seed envelopes} on the four headline cells. Across five seeds, the headline wins preserve direction.}
\label{fig:rob_b}
\end{figure}

\emph{Threat 2: run-to-run stochasticity.} We re-run Pass~2 five times ($n{=}5$) on the four headline cells. The original single-seed wins fall inside the multi-seed envelope and preserve direction (Figure~\ref{fig:rob_b}).

\begin{figure}[h]
\centering
\includegraphics[width=0.7\linewidth]{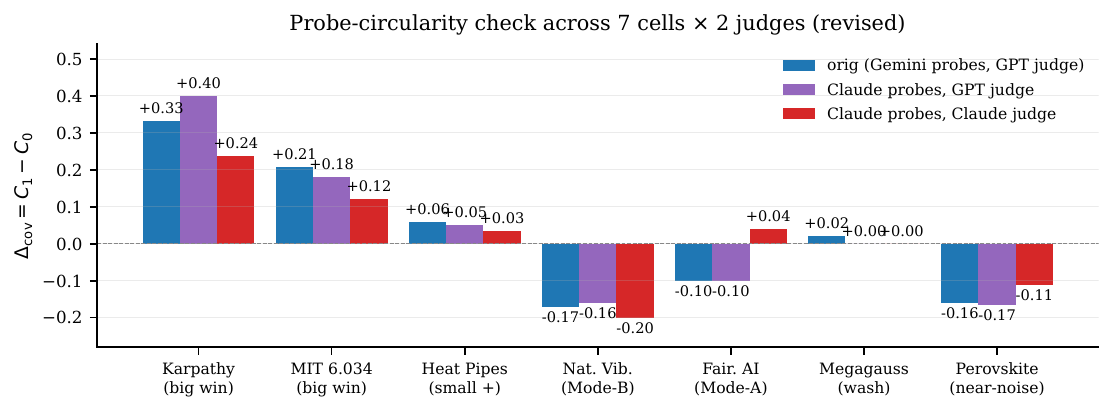}
\caption{\textbf{Threat 3 --- probe re-extraction} (7 cells). After Claude re-extracts the probes, the coverage direction is preserved on most cells under both judges.}
\label{fig:rob_c}
\end{figure}

\emph{Threat 3: probe circularity.} Because GPT-5.4 both extracts and scores the probes, we have a different model (Claude) re-extract the probes from the reference and re-score; the coverage direction is preserved on most cells under both judges (Figure~\ref{fig:rob_c}).

\begin{figure}[h]
\centering
\includegraphics[width=0.7\linewidth]{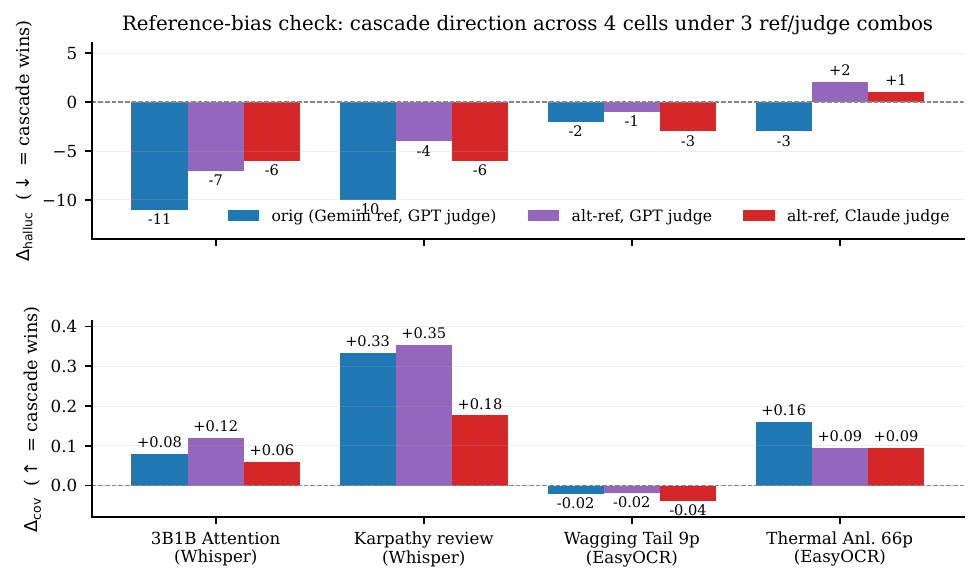}
\caption{\textbf{Threat 4 --- independent reference} (4 cells). An independent ASR/OCR reference preserves the coverage direction on all four cells.}
\label{fig:rob_d}
\end{figure}

\emph{Threat 4: shared-model reference bias.} Because Pass~1 and the reference transcript share a model, we regenerate the reference with an independent ASR/OCR system on four cells; the coverage direction survives on all four (Figure~\ref{fig:rob_d}). This control is small and concentrated on strong-win cases, so it remains the principal residual threat to the headline number.

\section{Multi-seed envelopes on the four headline cells}
\label{appdx:multiseed}

Figure~\ref{fig:rob_b} plots the envelopes; Table~\ref{tab:headline_multiseed_appdx} gives the numbers.

\begin{table}[h]
\centering
\small
\caption{Per-cell multi-seed envelopes for the four headline cells (Gemini $C_1$ Pass-2 at $T{=}0.7$, seeds 1--5). ``base-in-envelope?'' checks whether the $n{=}1$ value sits within the per-seed range.}
\label{tab:headline_multiseed_appdx}
\begin{tabular}{lrrrl}
\toprule
Cell & orig $n{=}1$ & $n{=}5$ halluc & $n{=}5$ cov & base-in-envelope? \\
\midrule
MIT 6.034 Winston   & $4$h / $0.98$ & $7.4 \pm 1.5$ & $0.971 \pm 0.028$ & cov yes; halluc \textbf{NO} (low) \\
Sapolsky Behavioral & $15$h / $0.94$ & $14.5 \pm 2.5$ & $0.898 \pm 0.017$ & halluc yes; cov \textbf{NO} (high) \\
Heat Pipes Gemini   & $11$h / $0.78$ & $15.4 \pm 2.7$ & $0.732 \pm 0.056$ & cov yes; halluc \textbf{NO} (low) \\
arXiv Perovskite    & $3$h / $0.74$ & $4.8 \pm 2.2$ & $0.844 \pm 0.056$ & halluc yes; cov \textbf{NO} (low) \\
\bottomrule
\end{tabular}
\end{table}

Two readings: (1)~the published $T{=}0$ baselines are mostly representative of the $T{=}0.7$ distribution on the headline cells, since every cell's $n{=}5$ envelope overlaps the published number on at least one axis. (2)~Perovskite was the seed-envelope point farthest from the multi-seed mean ($10$pp coverage below it), and under multi-seed verification its cascade-cov loss reads $-6$pp rather than $-16$pp. We retain Perovskite as a near-noise-floor cell and treat its earlier ``Mode-A counter-cell'' label as overstated at $n{=}1$.

\section{Three sources of stochasticity}
\label{appdx:stoch}

\begin{figure}[h]
\centering
\includegraphics[width=\linewidth]{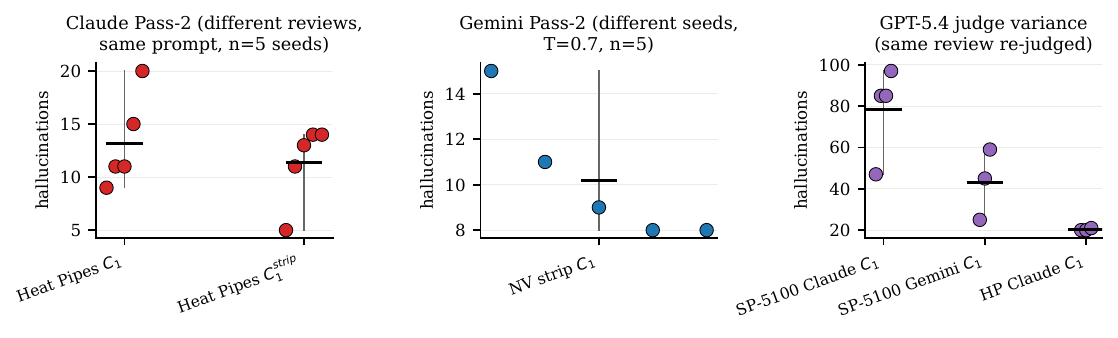}
\caption{\textbf{Three sources of stochasticity.} Left and middle: Pass-2 generation variance for Claude and Gemini. Right: GPT-5.4 re-judging the same review. Unsupported-claim counts vary much more than coverage, particularly for the high-count SP-5100 cell.}
\label{fig:stoch}
\end{figure}

\begin{figure}[h]
\centering
\includegraphics[width=\linewidth]{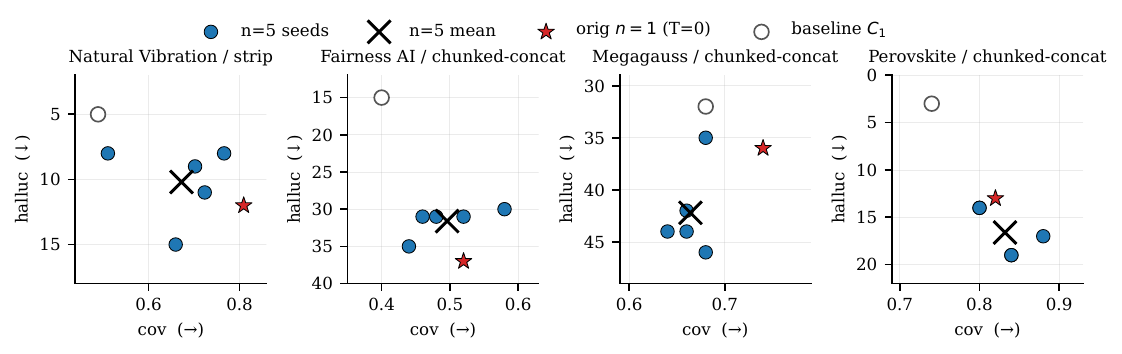}
\caption{\textbf{$n{=}5$ multi-seed envelopes for four intervention checks.} Blue dots are Pass-2 seeds ($T{=}0.7$), black $\times$ their mean, red stars the original $n{=}1$ result, and open circles plain $C_1$. The original result lies within the envelope except for Megagauss, whose chunked-Pass-2 coverage claim is therefore retracted.}
\label{fig:multiseed_envelopes_appdx}
\end{figure}

\paragraph{Detailed decomposition.} Figure~\ref{fig:stoch} plots the three sources; the per-cell breakdown follows in Figure~\ref{fig:multiseed_envelopes_appdx} and Table~\ref{tab:claude_stochasticity_appdx}.
\emph{(a)~Claude Pass-2 generation variance.} Same input + same API + same prompt: Heat~Pipes Claude $C_1$ swings $20\to 9$ between two runs (strip $13\to 14$ more stable), and the deltas are large enough that the strip effect cannot be distinguished from Pass-2 noise. Claude Opus 4.7 deprecates the \texttt{temperature} parameter so determinism cannot be enforced from the SDK. \emph{(b)~Gemini Pass-2 generation variance.} At \texttt{temperature=0.0}, two of three Gemini Pass-2 re-runs produced byte-identical output (MD5 match) on Heat~Pipes / SP-5100, and the third differed in similar length (1929 vs 1815 words, $\Delta h = +6$). Gemini is therefore mostly but not strictly deterministic, consistent with GPU-level non-associativity. \emph{(c)~GPT-5.4 high-reasoning judge variance.} Essentially deterministic on low-count cells ($\pm 1$ on Heat~Pipes Claude across three re-judgments) but substantially stochastic on high-count cells ($\pm 25$ on SP-5100, $\pm 17$ on the byte-identical SP-5100 Gemini review). We attribute this to long unsupported-claim lists running up against our 24k-token completion budget, so the judge truncates and resamples across calls. Coverage is stable on every cell across judge re-runs.

\paragraph{$n{=}5$ multi-seed verification on Heat Pipes Claude.} $C_1$ halluc $\{9,11,11,15,20\}$ (mean $13.2$, stdev $4.4$, range 9--20; a separate $n{=}5$ seed batch from the $14.8{\pm}3.5$ in Table~\ref{tab:iter_results}, both high-variance draws), and $C_1^{\rm strip}$ halluc $\{5,11,13,14,14\}$ (mean $11.4$, stdev $3.8$, range 5--14). Coverage means identical ($0.936$ both) with similar spread. Strip effect on hallucinations at $n{=}5$: $\hat{\Delta}=-1.8$, SE $\approx 2.6$, 95\% CI $[-6.9, +3.3]$, which crosses zero. \textbf{The Mode~B fix is statistically null on Claude/Heat~Pipes at $n{=}5$.}

\begin{table}[h]
\centering
\caption{Stochasticity decomposition. Top: Claude Pass-2 generation variance (different reviews, same prompt, judged once each). Bottom: GPT-5.4 judge variance on the \emph{same} Claude $C_1$ review file (one review, judged multiple times).}
\label{tab:claude_stochasticity_appdx}
\small
\begin{tabular}{lllllr}
\toprule
Cell & Condition & Run 1 & Run 2 & $\Delta h$ & $\Delta \mathrm{cov}$ \\
\midrule
\multicolumn{6}{l}{\emph{Claude Pass-2 generation variance}} \\
Heat Pipes 104p     & $C_1$           & 20h / 0.90 & 9h  / 0.90 & $-11$ & $\pm 0.00$ \\
                    & $C_1^{\rm strip}$ & 13h / 0.98 & 14h / 0.88 & $+1$  & $-0.10$ \\
SP-5100 176p        & $C_1$           & 47h / 0.81 & 79h / 0.81 & $+32$ & $\pm 0.00$ \\
                    & $C_1^{\rm strip}$ & 24h / 0.71 & 97h / 0.77 & $+73$ & $+0.06$ \\
Fairness AI 53p     & $C_1$           & 23h / 0.70 & 27h / 0.66 & $+4$  & $-0.04$ \\
                    & $C_1^{\rm strip}$ & 41h / 0.62 & 17h / 0.60 & $-24$ & $-0.02$ \\
\midrule
\multicolumn{6}{l}{\emph{GPT-5.4 judge variance (same review, judged multiple times)}} \\
Heat Pipes 104p     & $C_1$ (calls 1--3) & 20h / 0.90 & 21h / 0.92 & $+1$ & $+0.02$ \\
SP-5100 176p        & $C_1$ (calls 1--4) & 47h / 0.81 & 97h / 0.81 & $+50$ & $\pm 0.00$ \\
\bottomrule
\end{tabular}
\end{table}

\section{Inter-judge details}
\label{appdx:inter_judge}

Figure~\ref{fig:rob_a} plots the 22 cell--condition pairs (21 Gemini cells plus the Thermal-Claude cross-vendor cell); the tables below (Tables~\ref{tab:inter_judge_halluc} and~\ref{tab:inter_judge_cov}) give the per-cell counts on the original 10 cells (per-cell deltas for the remaining cells are released in the repository).

\begin{table}[h]
\centering
\small
\caption{Inter-judge hallucination counts on the original 10 cells. Direction $\checkmark$ means both judges agree on the sign of $C_0\to C_1$.}
\label{tab:inter_judge_halluc}
\begin{tabular}{lrrrrc}
\toprule
& \multicolumn{2}{c}{GPT-5.4} & \multicolumn{2}{c}{Claude} & Direction \\
\cmidrule(lr){2-3}\cmidrule(lr){4-5}
Cell & $C_0$ & $C_1$ & $C_0$ & $C_1$ & agree? \\
\midrule
Karpathy review            & 13 & 3  & 7  & 4  & \checkmark \\
Karpathy mtg-min           & 4  & 2  & 1  & 0  & \checkmark \\
SNAC mtg-min               & 14 & 11 & 10 & 11 & \texttimes  \\
Wagging Tail         & 3  & 1  & 2  & 2  & \texttimes  \\
Env Test            & 5  & 4  & 0  & 2  & \texttimes  \\
Dynamic Response    & 3  & 2  & 1  & 0  & \checkmark \\
Natural Vibration   & 4  & 5  & 2  & 4  & \checkmark \\
Thermal Analysis    & 5  & 2  & 2  & 1  & \checkmark \\
Heat Pipes         & 14 & 11 & 12 & 11 & \checkmark \\
Thermal Analysis Claude    & 11 & 4  & 9  & 4  & \checkmark \\
\bottomrule
\end{tabular}
\end{table}

\begin{table}[h]
\centering
\small
\caption{Inter-judge probe-coverage counts (probes covered out of $n$) on the original 10 cells.}
\label{tab:inter_judge_cov}
\begin{tabular}{lrrrrrc}
\toprule
& & \multicolumn{2}{c}{GPT-5.4} & \multicolumn{2}{c}{Claude} & Direction \\
\cmidrule(lr){3-4}\cmidrule(lr){5-6}
Cell & $n$ & $C_0$ & $C_1$ & $C_0$ & $C_1$ & agree? \\
\midrule
Karpathy review            & 51 & 32 & 49 & 38 & 49 & \checkmark \\
Karpathy mtg-min           & 50 & 8  & 17 & 13 & 21 & \checkmark \\
SNAC mtg-min               & 49 & 14 & 26 & 15 & 28 & \checkmark \\
Wagging Tail         & 49 & 39 & 38 & 40 & 38 & \checkmark \\
Env Test            & 50 & 44 & 46 & 46 & 49 & \checkmark \\
Dynamic Response    & 50 & 29 & 35 & 32 & 36 & \checkmark \\
Natural Vibration   & 47 & 31 & 23 & 32 & 28 & \checkmark \\
Thermal Analysis    & 50 & 34 & 42 & 34 & 45 & \checkmark \\
Heat Pipes         & 50 & 36 & 39 & 39 & 41 & \checkmark \\
Thermal Analysis Claude    & 50 & 46 & 50 & 46 & 49 & \checkmark \\
\bottomrule
\end{tabular}
\end{table}

Per-probe agreement (the fraction of probes the two judges classify identically as COVERED / MISSING) ranges $0.882$--$1.000$ across the (cell, condition) pairs, median $\approx 0.94$. Two of the three hallucination-direction disagreements on the original 10 cells (Wagging~Tail, Env~Test) are near-tie cells with $\le 5$ hallucinations under both judges. The third (SNAC) is driven by the judges assigning different $C_0$ baselines (14 vs 10) while agreeing on $C_1$.

\section{Probe-circularity and reference-bias check tables}
\label{appdx:probe_circ}

Figures~\ref{fig:rob_c} and~\ref{fig:rob_d} plot both checks; the tables below (Tables~\ref{tab:probe_circularity_appdx} and~\ref{tab:ref_bias_appdx}) give the numbers.

\begin{table}[h]
\centering
\small
\caption{Probe-circularity check on 7 cells spanning the cascade-direction spectrum. ``Original'' = published numbers (GPT-5.4-extracted probes, GPT-5.4 judge); the other columns re-extract probes via Claude and score under each judge.}
\label{tab:probe_circularity_appdx}
\begin{tabular}{lrrr}
\toprule
Cell (orig $\to$ Claude-extracted probes) & $\Delta_{\rm cov}$ (orig) & Claude-pr.\ / GPT & Claude-pr.\ / Claude \\
\midrule
Karpathy review (51 $\to$ 55) [audio]     & $+0.333$ & $+0.400$ & $+0.236$ \\
MIT 6.034 (48 $\to$ 50) [audio]            & $+0.208$ & $+0.180$ & $+0.120$ \\
Heat Pipes (50 $\to$ 60) [paper]     & $+0.060$ & $+0.050$ & $+0.033$ \\
Natural Vibration (47 $\to$ 50) [paper] & $-0.170$ & $-0.160$ & $-0.200$ \\
arXiv Fairness AI (50 $\to$ 50) [paper] & $-0.100$ & $-0.100$ & $+0.040$ \\
arXiv Megagauss (50 $\to$ 60) [paper]   & $+0.020$ & $+0.000$ & $+0.000$ \\
arXiv Perovskite (50 $\to$ 54) [paper]  & $-0.160$ & $-0.167$ & $-0.111$ \\
\bottomrule
\end{tabular}
\end{table}

\begin{table}[h]
\centering
\small
\caption{Reference-bias check on 4 cells. The non-Gemini reference rows replace OCR/ASR only, and probes are re-extracted by Claude. ``orig'' = Gemini reference + GPT-5.4 probes + GPT-5.4 judge.}
\label{tab:ref_bias_appdx}
\begin{tabular}{lcrrrr}
\toprule
Cell & Reference & $\Delta_h$ orig & $\Delta_h$ alt & $\Delta_{\rm cov}$ orig & $\Delta_{\rm cov}$ alt \\
\midrule
3B1B Attention 26m   & Whisper-base / GPT-5.4 & $-11$ & $-7$ & $+0.080$ & $+0.120$ \\
3B1B Attention 26m   & Whisper-base / Claude  & $-11$ & $-6$ & $+0.080$ & $+0.060$ \\
Karpathy review 42m  & Whisper-base / GPT-5.4 & $-10$ & $-4$ & $+0.333$ & $+0.353$ \\
Karpathy review 42m  & Whisper-base / Claude  & $-10$ & $-6$ & $+0.333$ & $+0.176$ \\
Wagging Tail      & EasyOCR / GPT-5.4      & $-2$  & $-1$ & $-0.020$ & $-0.019$ \\
Wagging Tail      & EasyOCR / Claude       & $-2$  & $-3$ & $-0.020$ & $-0.038$ \\
Thermal Analysis & EasyOCR / GPT-5.4      & $-3$  & $\mathbf{+2}$ & $+0.160$ & $+0.094$ \\
Thermal Analysis & EasyOCR / Claude       & $-3$  & $\mathbf{+1}$ & $+0.160$ & $+0.094$ \\
\bottomrule
\end{tabular}
\end{table}

The Thermal hallucination-direction flip under EasyOCR is a reference-side artifact: EasyOCR's transcript is $\approx 21\%$ shorter and loses figure captions, so claims the cascade correctly reproduces read as unsupported against the noisier reference.

\section{Per-cell hallucinations per 1k output words}
\label{appdx:perword}

\begin{table}[h]
\centering
\small
\caption{Hallucination density (per 1k output words) on Thermal Analysis.}
\label{tab:per_word_halluc}
\begin{tabular}{lrr}
\toprule
& $C_0$ & $C_1$ \\
\midrule
Gemini 3.1 Pro  & 4.2 & \textbf{1.4} \\
Claude Opus 4.7 & 3.7 & \textbf{1.4} \\
Mimo v2-omni    & 4.9 & 5.4 \\
\bottomrule
\end{tabular}
\end{table}

End-to-end hallucination rates differ per vendor (3.7 / 4.2 / 4.9 per 1k), and cascade brings Gemini and Claude to a common $\approx 1.4$ per 1k floor on the moderate-length paper (Table~\ref{tab:per_word_halluc}). Mimo does not converge to this floor.

\begin{table}[h]
\centering
\small
\caption{Hallucination density (per 1k output words) on Heat Pipes.}
\label{tab:per_word_halluc_hp}
\begin{tabular}{lrr}
\toprule
& $C_0$ & $C_1$ \\
\midrule
Gemini 3.1 Pro                            & 7.6 & 6.1 \\
Claude Opus 4.7                           & \textbf{1.2} & \textbf{9.6} \\
Claude Opus 4.7 ($C_1^{\rm strip}$)       & --- & 5.3 \\
Mimo v2-omni                              & 7.0 & 6.9 \\
\bottomrule
\end{tabular}
\end{table}

Claude's per-1k hallucination rate jumps from 1.2 ($C_0$) to 9.6 ($C_1$) on Heat~Pipes (Table~\ref{tab:per_word_halluc_hp}), an $8\times$ increase that survives length normalization. The same model on the same protocol on Thermal (66~pages) moved 3.7$\to$1.4. Citation stripping pulls Heat~Pipes Claude back to 5.3/1k. Gemini and Mimo on the same paper sit at $6$--$8$/1k regardless of cascade variant, indicating that the 1.4/1k floor reached on Thermal is not a property of the cascade in general, but of the specific (model, source) interaction.

\section{Specific hallucination examples (Heat Pipes Claude)}
\label{appdx:examples}

The 20 unsupported claims from Claude $C_1$ on Heat~Pipes (104~pages) are dominated by training-prior expansions of bibliographic abbreviations:

\begin{itemize}
\itemsep0pt
\item ``Cotter's 1965 analysis (LA-3246-MS)'': the transcript cites only ``Reference 2 (Cotter)'', so the year and report number are training-data details.
\item ``correlation from Allingham and McEntire (1961)'': the transcript only says ``Ref.~12'', so the author names and year are training-data details.
\item ``McAdams' laminar horizontal-tube formula'': the transcript says ``Reference 13 for laminar condensation on horizontal tubes'', so ``McAdams'' is a training-data detail.
\item ``Vapor chamber-fin radiator (Haller, Lieblein, and Lindow at NASA Lewis)'': the transcript says only ``proposed by Haller, et al of NASA''.
\end{itemize}

One outright misreading: the transcript states the no-boiling condition as $p_v - p_\ell \le 2\sigma\cos\theta/r_c$, but Claude's review writes $\ge$. This is a sign error in Pass-2 reading of Pass-1 text, not training-prior leak.

\section{Code and data availability}
\label{sec:code}
\label{appdx:code}

\begin{sloppypar}
All code, prompts, per-cell judgments, and the mechanistic-experiment artifacts (including the frontier modality probes, the running-example run directory, and the within-Claude cross-vendor comparison) are released in the project repository, whose README documents the pipelines and the exact commands to reproduce every cell, table, and figure in this paper.
\end{sloppypar}

\end{document}